\documentclass[10pt,journal]{IEEEtran}

\ifCLASSOPTIONcompsoc
  \usepackage[nocompress]{cite}
\else
  \usepackage{cite}
\fi
\ifCLASSINFOpdf
\else
\fi
\usepackage{amsmath}
\usepackage{times}
\usepackage{bm}
\usepackage{amsmath}
\usepackage{amssymb,amsfonts,graphicx,float,multirow,enumitem, url}
\usepackage{bbm}
\usepackage{comment}
\usepackage{placeins}
\usepackage{graphicx,caption} 
\usepackage{blindtext} 
\usepackage{adjustbox} 

\newtheorem{theorem}{Theorem}

\newtheorem{remark}{Remark}

\def\P{{\textsf{pr}}}
\def\E{{\textsf{E}}}
\def\Var{{\textsf{var}}}
\def\Cov{{\textsf{cov}}}
\def\Sig{\Sigma}
\allowdisplaybreaks
\def\B{\bf}
\def\c{\centering}

\newcommand{\pr}{\textsf{pr}} 
\newcommand{\ep}{\textsf{E}} 
\newcommand{\var}{\textsf{var}} 
\newcommand{\cov}{\textsf{cov}} 

\usepackage{color} 
\definecolor{red2}{rgb}{0.7, 0, 0.1}

\usepackage{soul} 

\begin{document}
%
\title{A Fast and Efficient Change-point Detection Framework based on Approximate $k$-Nearest Neighbor Graphs}
%
%
%
%

\author{Yi-Wei~Liu
        and~Hao~Chen

\IEEEcompsocitemizethanks{\IEEEcompsocthanksitem Y. Liu (lywliu@ucdavis.edu) and H. Chen (hxchen@ucdavis.edu) are with the Department of Statistics, University of California, Davis, Davis, CA, 95616.  The authors are supported in part by the NSF awards DMS-1513653 and DMS-1848579.
\IEEEcompsocthanksitem }}

%
%

\markboth{ }%
{Y. Liu and H. Chen: Change-point Detection based on approximate $k$-NN Graphs}
%



\IEEEtitleabstractindextext{%
\begin{abstract}
Change-point analysis is thriving in this big data era to address problems arising in many fields where massive data sequences are collected to study complicated phenomena over time. It plays an important role in processing these data by segmenting a long sequence into homogeneous parts for follow-up studies. The task requires the method to be able to process large datasets quickly and deal with various types of changes for high-dimensional data. We propose a new approach making use of approximate $k$-nearest neighbor information from the observations, and derive an analytic formula to control the type I error. The time complexity of our proposed method is $O\left(dn(\log n+k \log d)+nk^2\right)$ for an $n$-length sequence of $d$-dimensional data. The test statistic we consider incorporates a useful pattern for moderate- to high- dimensional data so that the proposed method could detect various types of changes in the sequence.  The new approach is also asymptotic distribution free, facilitating its usage for a broader community.  We apply our method to fMRI datasets and Neuropixels datasets to illustrate its effectiveness.
\end{abstract}

}

\maketitle

\IEEEdisplaynontitleabstractindextext

%
\IEEEpeerreviewmaketitle

\ifCLASSOPTIONcompsoc
\IEEEraisesectionheading{\section{Introduction}\label{sec:introduction}}
\else
\section{Introduction}
\label{sec:introduction}
\fi


\IEEEPARstart{W}{ITH} advances in technologies, scientists in many fields are collecting massive data for studying complex phenomena over time and/or space. Such data often involve sequences of high-dimensional measurements that cannot be analyzed through traditional approaches. Insights on such data often come from segmentation/change-point analysis, which divides the sequence into homogeneous temporal or spatial segments. They are crucial early steps in understanding the data and in detecting anomalous events. Change-point analysis has been extensively studied for univariate and low-dimensional data (see \cite{BN1993,BD1993,Carlstein1994,CH1997,ChenGupta} for various aspects of classic change-point analysis). However, many modern applications require effective and fast change-point detection for high-dimensional data. For example, Neuropixels recordings \cite{Jun2017}, microarrays \cite{Z2018microarrays}, healthcare data \cite{Lee2017healthcare}, etc.

Let the sequence of observations be $\{\mathbf{y}_t:t=1,\ldots,n\}$, indexed by some meaningful order, such as time or location. Then the change-point detection problem can be formulated as testing the null hypothesis of homogeneity:
\begin{eqnarray} \label{H0}
H_0: \mathbf{y}_t\sim F_0, \, t=1,\ldots,n,
\end{eqnarray}
against the alternative that there exists a change-point $\tau$:
\begin{eqnarray}
H_1: \exists 1\leq \tau<n, \, \mathbf{y}_t\sim \left\{
\begin{aligned}
F_0 & , \, \, t\leq \tau \\
F_1 & , \, \, \text{otherwise}
\end{aligned}
\right.
\end{eqnarray}
Here, $F_0$ and $F_1$ are two different probability measures. When there are multiple change-points, wild binary segmentation \cite{fryzlewicz2020detecting,fryzlewicz2014wild} or seeded binary segmentation \cite{kovacs2020seeded} can be incorporated.

Recently, there were quite a number of progresses on parametric change-point detection. For example, \cite{B2018TS} used the piecewise stationary time series factor models for multivariate observations, and assumed the changes are in their second-order structure. Reference \cite{Wang2018-1} studied the change-point detection and localization problem in dynamic networks by assuming the entries of the adjacency matrices are from inhomogeneous Bernoulli models. Reference \cite{B2018} considered the change-point detection problem in networks generated by a dynamic stochastic block model mechanism. Reference \cite{L2021Graph} addressed the change-point problem with missing values, and focused on detecting the covariance structure breaks in Gaussian graphical models. These methods work under certain parametric models. However, in many applications, we have little knowledge on $F_0$ and $F_1$.

In the context of nonparametric change-point detection for high-dimensional data, kernel-based methods were first explored \cite{H2007,H2009}, and continued to be improved \cite{H2019}. However, this kernel approach is difficult to use practically. As we will show in Section \ref{Section.4}, this method is very sensitive to the choice of a tuning parameter and it is time-consuming to apply the method with a proper type I error control, {{where type I error is the event that a change-point is falsely detected when the sequence is actually homogeneous.}} Reference \cite{Li2019scanB} proposed the scan $B$-statistic for kernel change-point detection, which is computationally efficient and has a fast formula for type I error control; however, it requires a large amount of reference data. In recent years, distance-based methods \cite{Matteson2014} and graph-based methods \cite{Chen2015,Chu2019} were proposed for high-dimensional change-point detection. The distance-based method (ecp) uses all pairwise distances among observations to find change-points, which could also be computationally heavy for large datasets because computing all pairwise distances needs $O(dn^2)$ time for $d$-dimensional data. In addition, there is no fast analytic formula for type I error control, and thus one needs to draw random permutations to approximate the $p$-value.
The graph-based methods \cite{Chen2015,Chu2019} utilize the information of a similarity graph constructed on observations to detect change-points. The authors also provided analytic formulas for type I error control, making them faster to run. In addition, the graph-based methods can detect more types of changes compared to ecp. {{The ecp method is very sensitive to changes in mean but its performance decays when the changes come in many other forms}} (see Section \ref{Power}).

In this paper, we seek further improvement on graph-based methods, especially from an efficiency perspective. Existing graph-based methods for offline change-point detection utilize an undirected graph constructed among observations. Some common choices are the minimum spanning tree (MST), where all observations are connected with the total distance minimized; the minimum distance pairing (MDP), where the observations are partitioned into $n/2$ pairs with the total within-pair distance minimized; the undirected nearest neighbor (NN) graph, where each observation connects to its nearest neighbor; and their denser versions, $k$-MST, $k$-MDP, and undirected $k$-NN graphs. Take the $k$-MST for example, it is the union of the $1$st, $\ldots$, $k$th MSTs, where the $1$st MST is the MST, and the $j$th MST is a spanning tree connecting all observations such that the sum of the edges in the tree is minimized under the constraint that it does not contain any edge in the $1$st, $\ldots$, $(j-1)$th MSTs. Among these graphs, $k$-MST is preferred as it in general has a higher power than others \cite{Chen2015}. Nevertheless, it requires $O(dn^2)$ time to compute the distance matrix among $n$ $d$-dimensional observations, so it takes at least $O(dn^2)$ time to construct the $k$-MST from the original data when the pairwise distances were not provided in the beginning, which is usually the case. This could be inefficient when either $n$ or $d$ is large.

Hence, we seek other ways to construct the similarity graph. There are fast existing algorithms to construct the directed approximate $k$-NN graph \cite{FNN}, where each observation finds $k$ other nearby points that might not be the $k$ closest ones.  {{We use the $kd$-tree algorithm to search for approximate nearest neighbors. A $kd$-tree is a space-partitioning data structure for organizing points in a high-dimensional space, which is a binary tree constructed through splitting the points by the values on alternating coordiantes as the tree grows.  It takes $O(dn\log n)$ time to preprocess a set of $n$ points in $\mathbb{R}^d$ \cite{arya1998optimal}. The nearest neighbor for any given query point can be searched efficiently with the $kd$-tree. To approximate the nearest neighbors, first traverse the tree to the leaf node that contains the query point, and then search for the nearest neighbors only in nearby areas. It requires only $O(d\log d+\log n)$ time to result in a good approximate nearest neighbor per query \cite{Ram2019}, so the total computational cost for obtaining a directed approximate $k$-NN graph with the $kd$-tree can be achieved at $O\left(dn(\log n+k\log d)\right)$. }}    Simulation studies show that this new approach has power on par with the existing method on $k$-MST (Section \ref{Power}), and the new approach is much faster (Section \ref{Time}).

Since the existing offline graph-based change-point detection framework needs the graph to be an undirected graph, we further work out a framework that can deal with the directed approximate $k$-NN graph, i.e., all the following steps after the graph is constructed: the exact analytic formulas to compute the test statistic, the limiting distribution of the new statistic, and the analytic formula 
to supervise the false discovery rate efficiently.  The time complexity of the method after the directed approximate $k$-NN graph is obtained is $O(nk^2)$. Thus, the overall time complexity of the new method is $O\left(dn(\log n+k\log d)+nk^2\right)$. We illustrate the new approach on the analyses of fMRI datasets and Neuropixels datasets (Section \ref{application}). The former ones have very large dimensions and moderate sample sizes, whereas the latter ones feature  very large sample sizes with moderate dimensions.


\section{Proposed Statistic}

\begin{figure}[h]
\center
\captionsetup{width=0.95\linewidth} 
\includegraphics[width=0.32\linewidth]{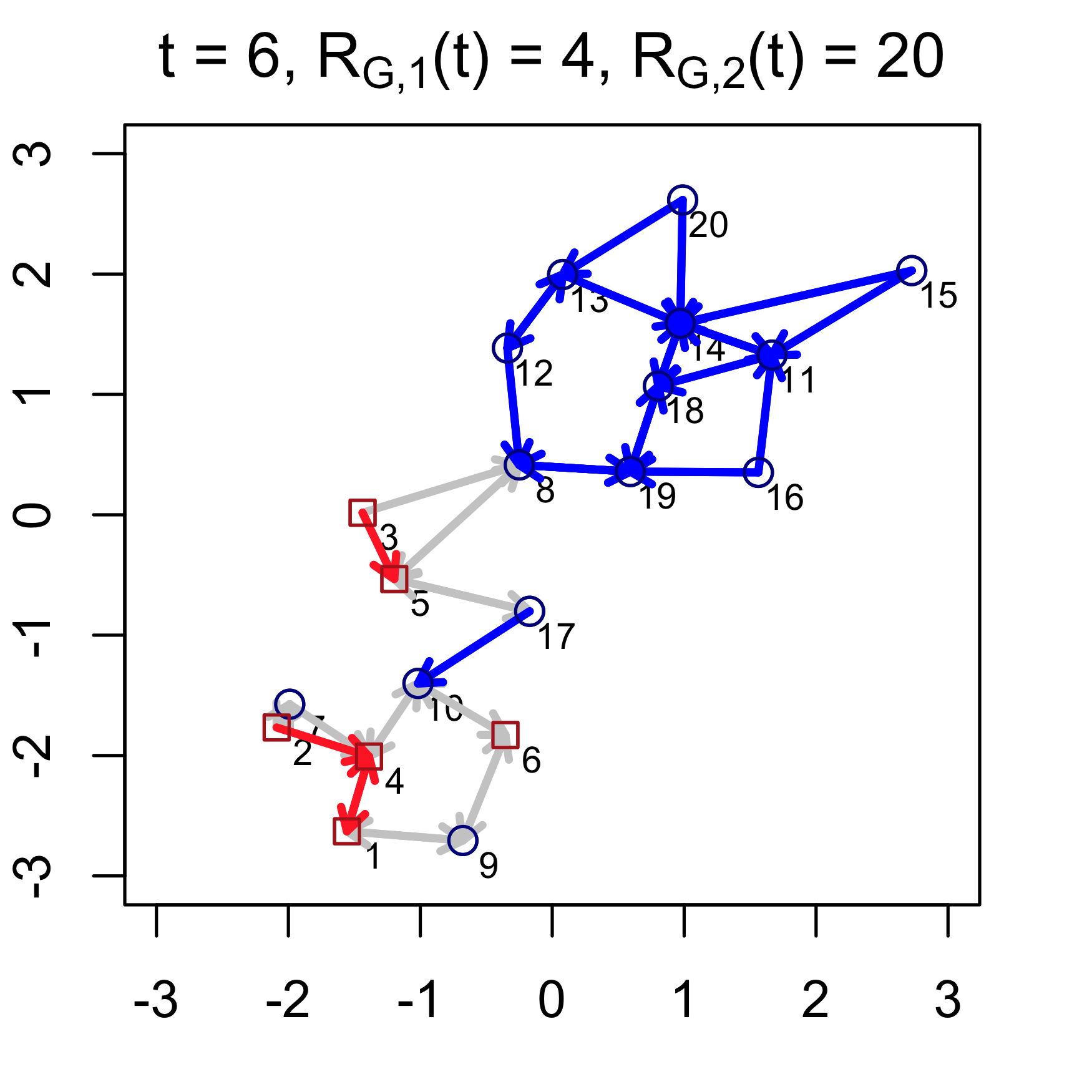}
\includegraphics[width=0.32\linewidth]{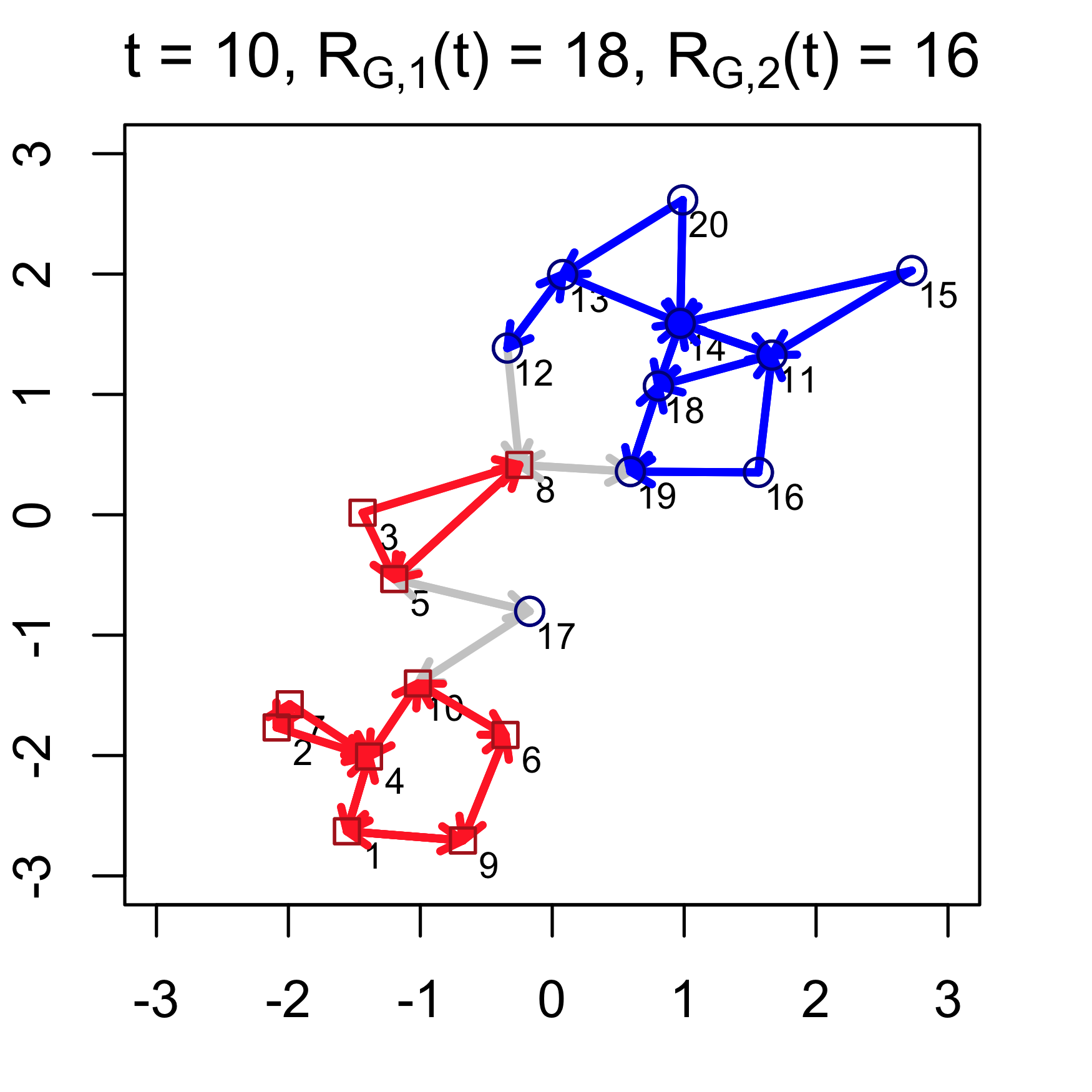}
\includegraphics[width=0.32\linewidth]{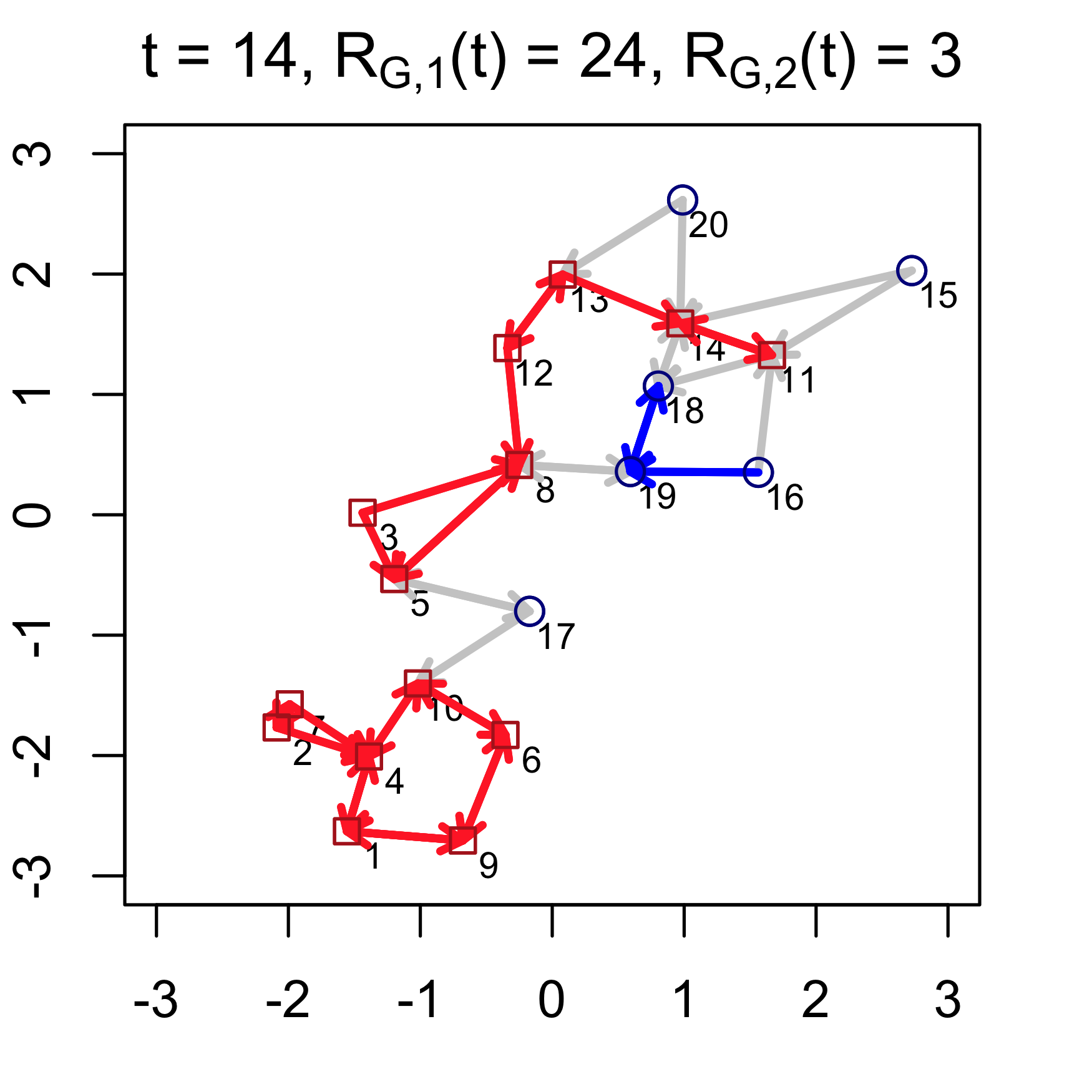}
\caption{The computation of $R_{G,1}(t)$ and $R_{G,2}(t)$ at three different values of $t$. Here $\mathbf{y}_1,\ldots,\mathbf{y}_{10} \stackrel{\text{i.i.d.}}{\sim} N((-0.5,-0.5)^T,\mathbbm{I}_2)$, and $\mathbf{y}_{11},\ldots,\mathbf{y}_{20} \stackrel{\text{i.i.d.}}{\sim} N((0.5,0.5)^T,\mathbbm{I}_2)$, where $\mathbbm{I}_2$ is the $2\times 2$ identity matrix. The graph $G$ here is the directed 2-NN on the Euclidean distance. Each $t$ divides the observations into two groups: one group for observations before $t$ (red squares) and the other group for observations after $t$ (blue circles). Red edges connect observations before $t$ and the number of red edges is $R_{G,1}(t)$; blue edges connect observations after $t$ and the number of blue edges is $R_{G,2}(t)$. Notice that as $t$ changes, the group identities change but the graph $G$ does not change.}
\label{graph}
\end{figure}

Let $G=\{(i,j): \mathbf{y}_j$  is among  $\mathbf{y}_i$'s $ k$  approximate nearest neighbors\} be the directed approximate $k$-NN graph, $R_{G,1}(t)$ be the number of edges on $G$ connecting observations both before $t$, and $R_{G,2}(t)$ be the number of edges on $G$ connecting observations both after $t$: $$R_{G,1}(t)=\sum_{(i,j)\in G} \mathbbm{1}_{\{i\leq t, j\leq t\}}, \ R_{G,2}(t)=\sum_{(i,j)\in G} \mathbbm{1}_{\{i> t, j> t\}};$$
with $\mathbbm{1}_{A}$ being the indicator function for event $A$. Here, we use the notations similar to those in \cite{Chu2019}. A key difference is that the graph in \cite{Chu2019} is undirected, whereas here $G$ is directed. Fig. \ref{graph} illustrates the computation of $R_{G,1}(t)$ and $R_{G,2}(t)$ on a toy example. We will use these two quantities to construct the test statistics. The rationale is as follows: When all observations are from the same distribution, the distributions of $R_{G,1}(t)$ and $R_{G,2}(t)$ can be figured out under the permutation null distribution that places $1/n!$ probability on each of the $n!$ permutations of $\{\mathbf{y}_i: i = 1,\dots,n\}$. 
With no further specification, we use $\pr$, $\ep$, $\var$, and $\cov$ to denote probability, expectation, variance, and covariance, respectively, under the permutation null distribution. When there is a change-point at $\tau$, one typical outcome is that observations from the same distribution tend to form edges within themselves, making both $R_{G,1}(\tau)$ and $R_{G,2}(\tau)$ larger than their null expectations. Another common but somewhat counter-intuitive outcome is that observations from one distribution tend to connect within themselves, but observations from the other distribution tend not to connect within themselves, casuing one of $R_{G,1}(\tau)$ and $R_{G,2}(\tau)$ to be larger than its null expectation, and the other smaller than its null expectation. This happens commonly under moderate to high dimensions when the variances of the two distributions differ. The underlying reason is the curse of dimensionality (see \cite{CF2017} for detailed explanations on this phenomenon under the two-sample testing setting).

To cover both possible outcomes under the alternative, we focus on a max-type test statistic in the main context. Three other test statistics (original/weighted/generalized) are discussed in Supplement D.

For each candidate $t$ of the true change-point $\tau$, the max-type edge-count statistic is defined as
\begin{eqnarray} \label{max-type}
M(t) = \max(Z_w(t),|Z_{\text{diff}}(t)|);
\end{eqnarray}
where
\begin{eqnarray*}
Z_w(t) &=& \frac{R_w(t)-\E(R_w(t))}{\sqrt{\Var(R_w(t))}}, \\
Z_{\text{diff}}(t) &= & \frac{R_{\text{diff}}(t)-\E(R_{\text{diff}}(t))}{\sqrt{\Var(R_{\text{diff}}(t))}};
\end{eqnarray*}
with
\begin{eqnarray*}
R_w(t) &=& \frac{n-t-1}{n-2}R_{G,1}(t)+\frac{t-1}{n-2}R_{G,2}(t), \\
R_{\text{diff}}(t) &=& R_{G,1}(t)-R_{G,2}(t).
\end{eqnarray*}
The null hypothesis of homogeneity (\ref{H0}) is rejected if the test statistic
\begin{eqnarray}
\max_{n_0\leq t\leq n_1}M(t);
\end{eqnarray}
with $n_0$ and $n_1$ pre-specified, is larger than the critical value for a given significance level, {{which measures the strength of the evidence that must be presented in the sample to reject the null hypothesis, and has to be determined before conducting the experiment. Statistically, significance level is the probability of rejecting the null hypothesis when it is true, usually set to be 5\% or 1\%.}}

\begin{figure}[h]
\hspace{-0.1cm} {\textbf{(a)}} \hspace{1.5cm} {\textbf{(b)}} \hspace{1.5cm} {\textbf{(c)}} \hspace{1.5cm} {\textbf{(d)}} 
\vspace{-0.55em}
\center
\includegraphics[width=0.24\linewidth]{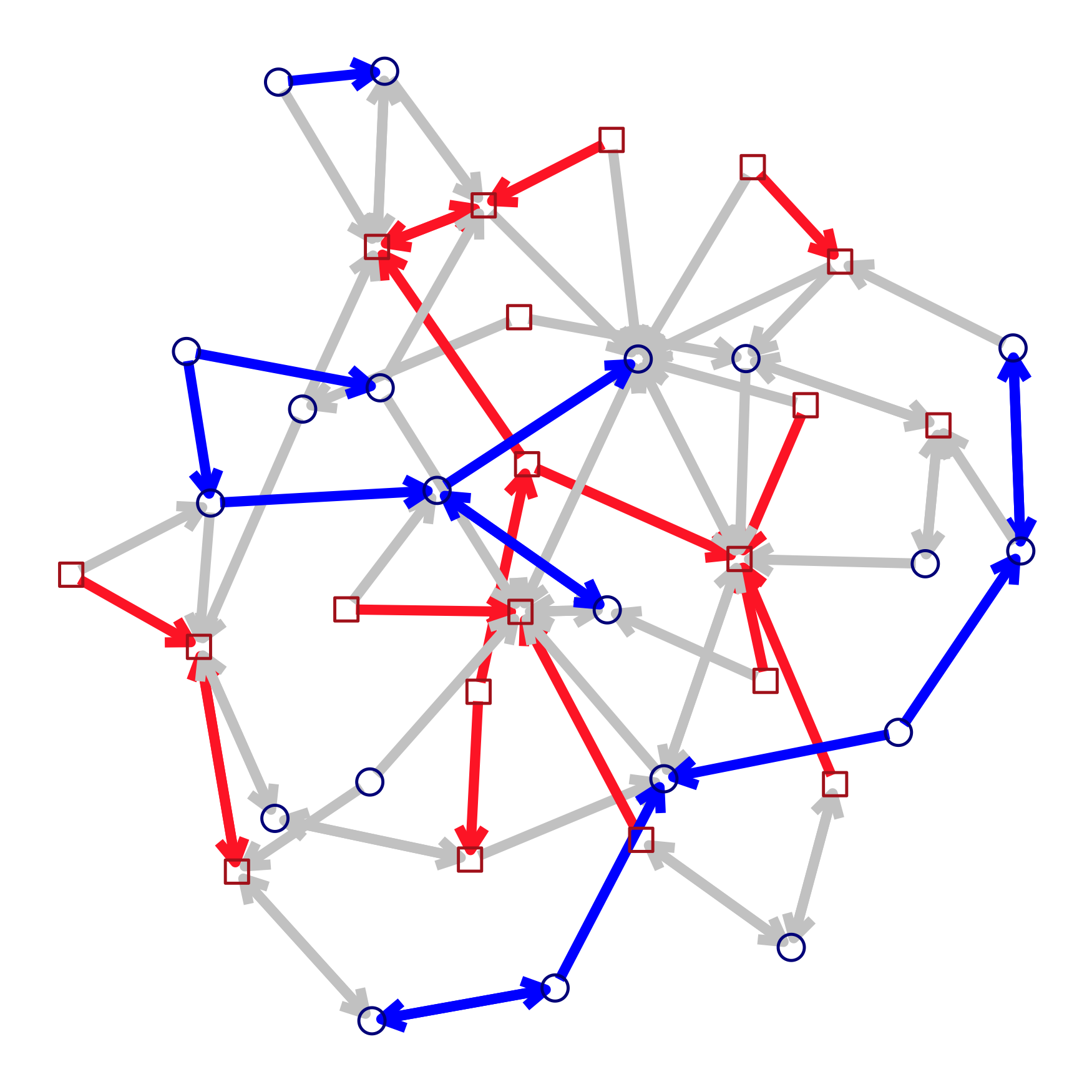}
\includegraphics[width=0.24\linewidth]{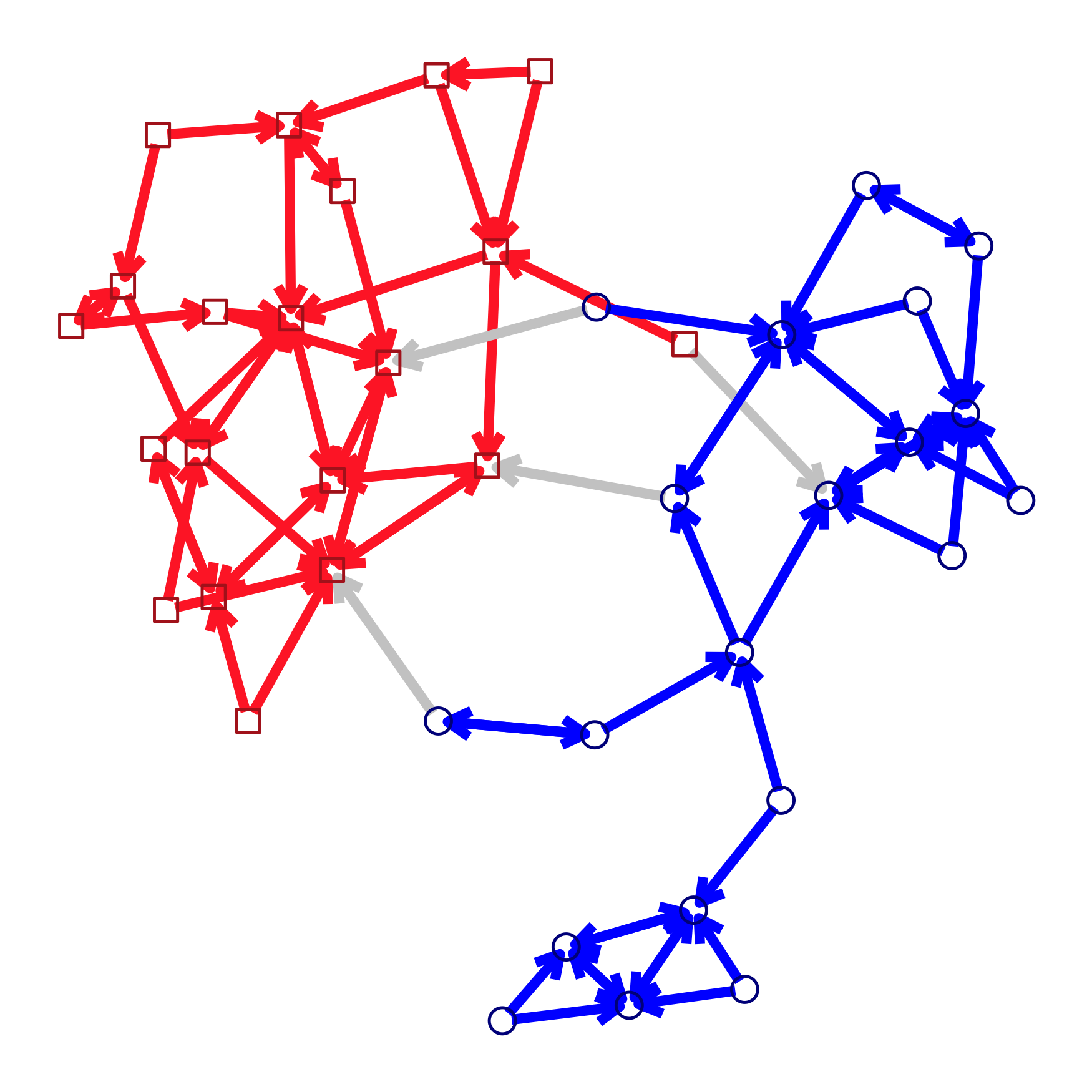}
\includegraphics[width=0.24\linewidth]{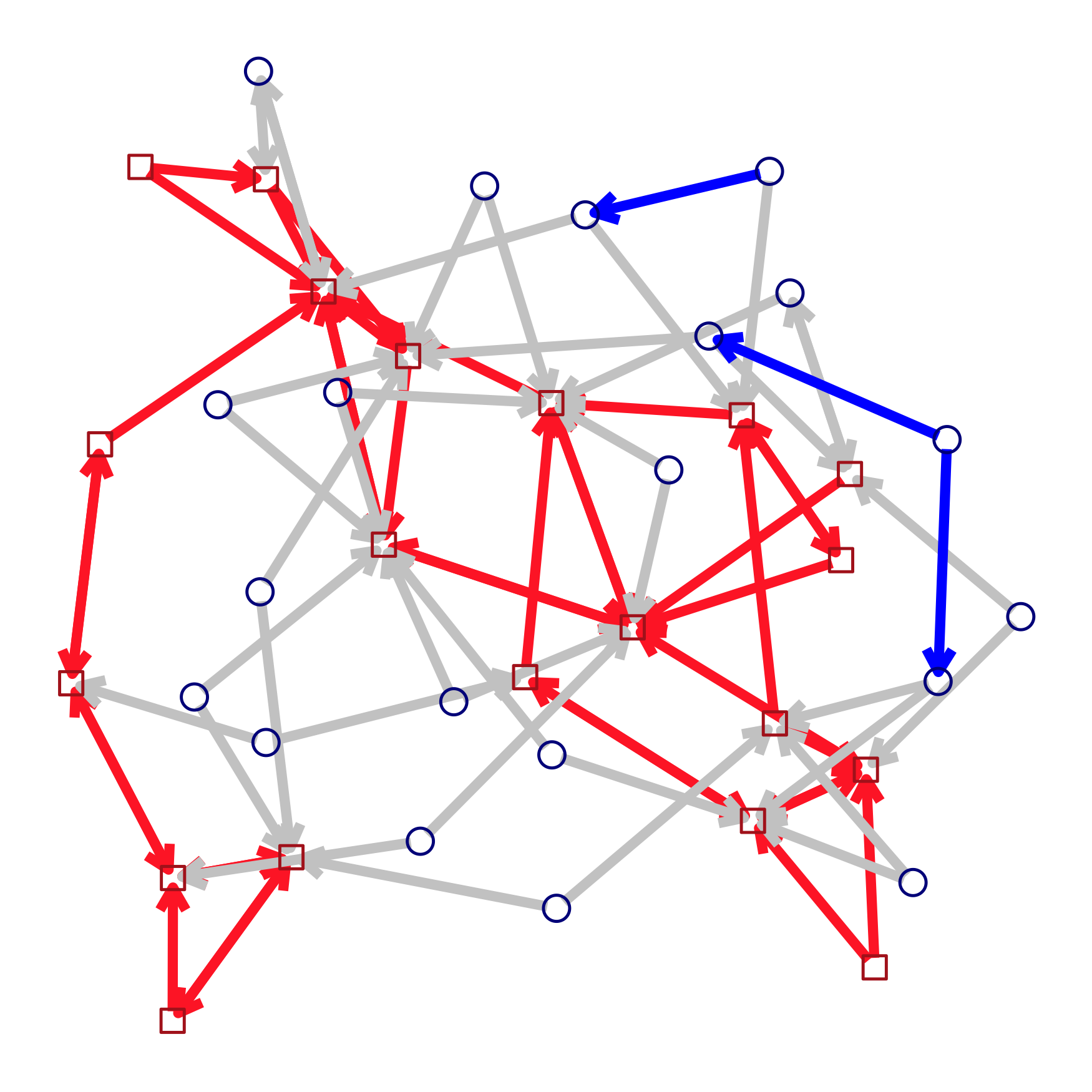}
\includegraphics[width=0.24\linewidth]{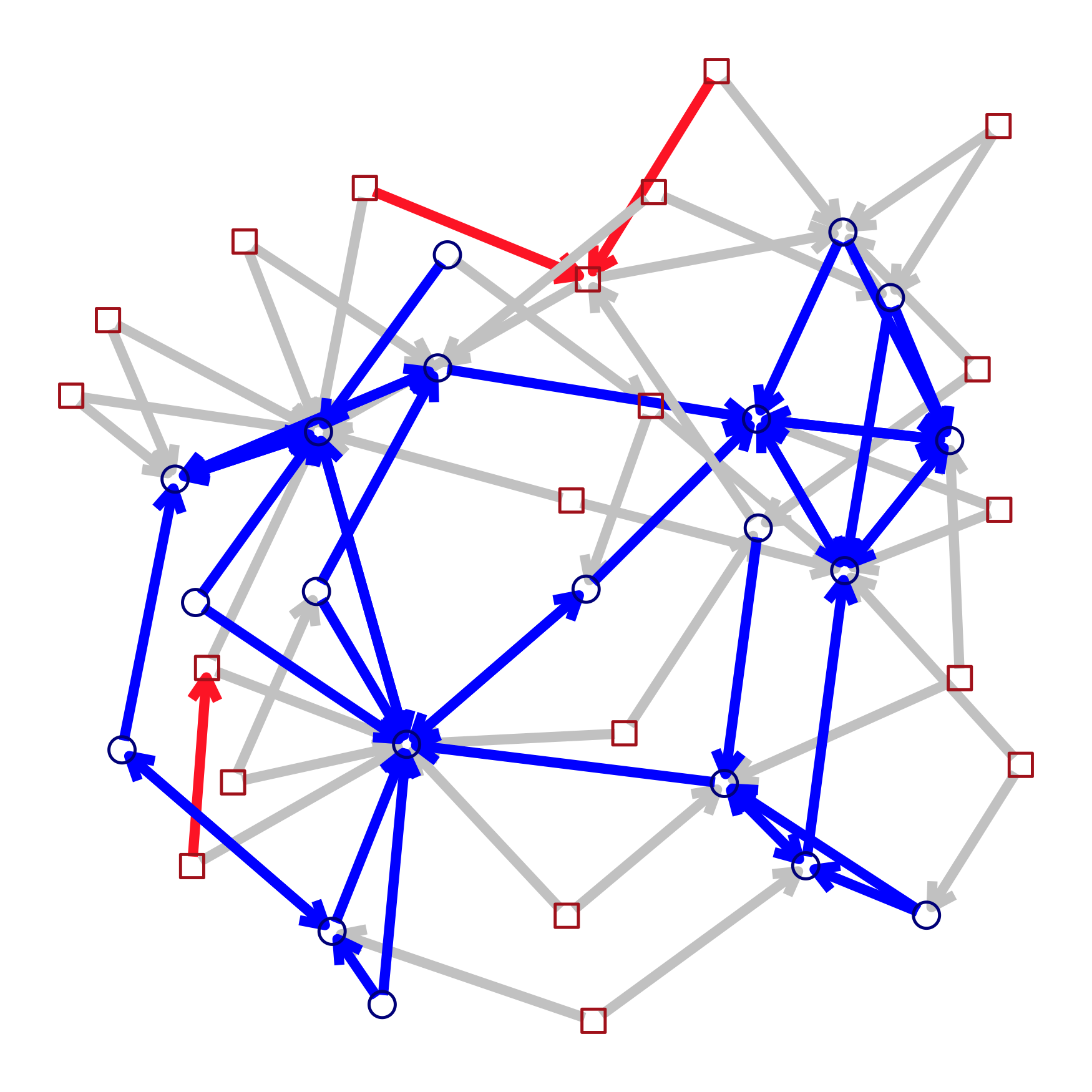}
\caption{Directed 2-NN graphs on 100-dimensional data visualized by the \texttt{ggnet2} function.  Here, $\mathbf{y}_1,\ldots,\mathbf{y}_{20}$ (red squares) are randomly drawn from a 100-dimensional Gaussian distribution with zero mean and identity covariance matrix, and $\mathbf{y}_{21},\ldots,\mathbf{y}_{40}$ (blue circles) are randomly drawn from $N_{100}(\bm{\mu}, a \mathbbm{I}_{100})$ with (a) $\bm{\mu}=0\times \mathbf{1}_{100}, a=1$; (b) $\bm{\mu}=0.8\times \mathbf{1}_{100}, a=1$; (c) $\bm{\mu}=0\times \mathbf{1}_{100}, a=1.4$; (d) $\bm{\mu}=0\times \mathbf{1}_{100}, a=0.8$, where $\mathbf{1}_{100}$ is a length-100 vector whose elements are all one's. The same edge coloring scheme as in Fig. \ref{graph} is used here.}
\label{picture}
\end{figure}

Here, the two components, $Z_w(t)$ and $|Z_{\text{diff}}(t)|$, capture the aforementioned two possible outcomes under the alternative. For better understanding, we illustrate the outcomes through a toy example (Fig. \ref{picture}). When $\{\mathbf{y}_1,\ldots,\mathbf{y}_t\}$ and $\{\mathbf{y}_{t+1},\ldots,\mathbf{y}_n\}$ are from the same distribution, they are well mixed and $R_{G,1}(t)$ and $R_{G,2}(t)$ would be close to their null expectations (Fig. \ref{picture} (a)). When they are from different distributions, one common exhibition is that observations from the same distribution are more likely to be connected in $G$. Fig. \ref{picture} (b) plots a typical directed $2$-NN graph under this alternative and we see that there are more edges connecting within each group. When this happens, $Z_w(t)$ is large. For moderate- to high- dimensional data, another exhibition of the graph is common under the alternative shown in Fig. \ref{picture} (c). Here, the dimension is $d=100$, and the blue circles are from a distribution with a larger variance than that of the red squares. We see that $R_{G,1}(t)$ is much larger than its null expectation but $R_{G,2}(t)$ is much smaller than its null expectation (very few blue edges). This happens due to the curse of dimensionality. As the volume of a $d$-dimensional ball increases exponentially in $d$, the blue circles from a distribution with a larger variance are sparsely scattered and tend to find their nearest neighbors in red squares. The $Z_{\text{diff}}(t)$ part in our statistic is effective in capturing this pattern. The absolute value is to cover the two possible scenarios in opposite directions showcased in Fig. \ref{picture} (c) and (d). 

We next provide the exact analytic formulas for the expectation and variance of $\left(R_{G,1}(t),R_{G,2}(t)\right)^T$ that are required to compute $M(t)$ so that we do not need to perform the time-consuming permutations to obtain them.

\begin{theorem} 
\label{theorem1}
The expectation, variance, and covariance of $R_{G,1}(t)$ and $R_{G,2}(t)$ under the permutation null distribution are:
\begin{flalign*} 
 & \E\left(R_{G,1}(t)\right) = nkp_1(t), \quad \E\left(R_{G,2}(t)\right) = nkq_1(t), & \\
 & \Var\left(R_{G,1}(t)\right) = d_1 p_1(t) + d_2 p_2(t) + d_3 p_3(t) - \left(nkp_1(t)\right)^2, & \\
 & \Var\left(R_{G,2}(t)\right) = d_1 q_1(t) + d_2 q_2(t) + d_3 q_3(t) - \left(nkq_1(t)\right)^2, & \\
 & \Cov\left(R_{G,1}(t),R_{G,2}(t)\right) = d_3 r(t) - \left(nkp_1(t)\right)\left(nkq_1(t)\right), &
\end{flalign*}
where
\begin{flalign*}
& \quad p_1(t)=\frac{t(t-1)}{n(n-1)}, \quad p_2(t)=\frac{t(t-1)(t-2)}{n(n-1)(n-2)}, & \\
& \quad p_3(t)=\frac{t(t-1)(t-2)(t-3)}{n(n-1)(n-2)(n-3)}, & \\
& \quad q_1(t)=\frac{(n-t)(n-t-1)}{n(n-1)}, & \\
& \quad q_2(t)=\frac{(n-t)(n-t-1)(n-t-2)}{n(n-1)(n-2)}, & \\
& \quad q_3(t)=\frac{(n-t)(n-t-1)(n-t-2)(n-t-3)}{n(n-1)(n-2)(n-3)}, & \\
& \quad r(t)=\frac{t(t-1)(n-t)(n-t-1)}{n(n-1)(n-2)(n-3)}, &
\end{flalign*}
and $d_1=c^{(1)}+c^{(2)}$, $d_2=c^{(3)}+c^{(4)}+c^{(5)}+c^{(6)}$, $d_3=c^{(7)}$, where $c^{(1)},\ldots,c^{(7)}$ are quantities on the graph $G$, defined as:
\begin{flalign*}
& \quad c^{(1)} = nk, \quad c^{(2)} = \sum_{i=1}^n  \sum_{j\in D_i} \mathbbm{1}_{\{(i,j)\in G\}}, & \\
& \quad c^{(3)} = c^{(4)} =\sum_{i=1}^n  \sum_{j\in D_i} (k - \mathbbm{1}_{\{(i,j)\in G\}}), \quad c^{(5)} = nk(k-1), & \\
& \quad c^{(6)} = \sum_{i=1}^n  \left( |D_i|^2 - |D_i| \right), \quad c^{(7)} = (nk)^2 - \sum_{m=1}^6 c^{(m)}. &
\end{flalign*}
Here, $D_i$ is the set of indices of observations that point toward observation $\mathbf{y}_i$, and $|D_i|$ is the cardinality of set $D_i$, or the in-degree of observation $\mathbf{y}_i$.
\end{theorem}

\begin{remark}
The time complexity of computing $c^{(1)},\ldots,c^{(7)}$ in Theorem \ref{theorem1} is $O(nk)$. One only has to construct a list of in-degrees for each observation in order to compute these seven quantities, which takes $O(nk)$ time.
\end{remark}

\begin{figure}[h]
\center
\includegraphics[width=1.0\linewidth]{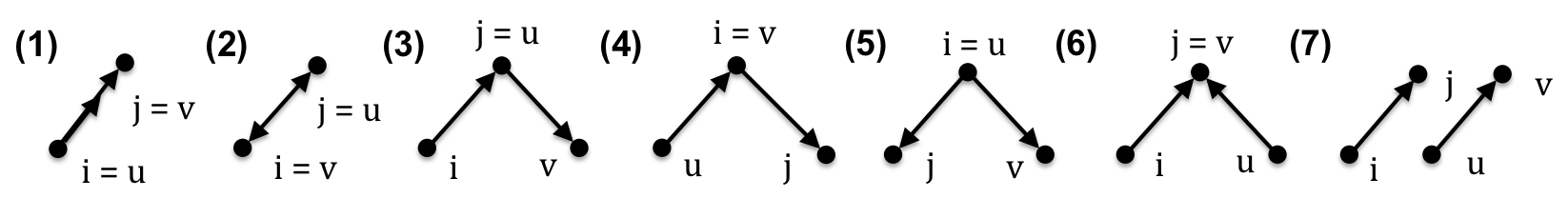}
\caption{Seven possible configurations of two edges $(i,j), (u,v)$ randomly chosen with replacement from a directed graph: (1) two edges degenerate into one ($(i,j)=(u,v)$); (2) two opposite edges (the two end nodes point to each other); (3)-(6) four different configurations with the two edges sharing one node; (7) two edges without any node sharing.}
\label{figure1}
\end{figure}

Theorem \ref{theorem1} can be proved by combinatorial analysis. The expectations can be obtained easily by the linearity of expectation. For the variances and the covariance, we have to figure out the numbers of the seven possible configurations of pairs of edges as plotted in Fig. \ref{figure1}. The quantities $c^{(1)},\ldots,c^{(7)}$ in Theorem \ref{theorem1} correspond respectively to the numbers of the seven configurations on a directed approximate $k$-NN graph. For such graphs, the out-degree of every observation node is a constant $k$, while the in-degree could vary from node to node, which requires one to scan through every edge to obtain the information.
A detailed proof of Theorem \ref{theorem1} is in Supplement A.1.

This max-type edge-count statistic $M(t)$ in (\ref{max-type}) is well-defined under very mild conditions (Theorem \ref{theorem2}). The proof is in Supplement A.2.

\begin{theorem} \label{theorem2}
The max-type edge-count statistic $\{M(t)\}_{t=1,\dots,n-1}$ on a directed approximate $k$-NN graph is well-defined when $n\geq 5$ and not every observation has the same in-degree (i.e., there exists an $i$, $1\leq i \leq n$, such that $|D_i|\neq k$).
\end{theorem}

The conditions in Theorem \ref{theorem2} ensure that the variances of $R_w(t)$ and $R_{\text{diff}}(t)$ are not zero. If all the observations have the same in-degree $k$, then $R_{\text{diff}}(t)$ is a constant. As the distribution of in-degrees could vary, we examine the most extreme case of which on the directed $k$-NN graph. Under the worst scenario, $n\geq 5$ ensures that the variance of $R_w(t)$ is positive.


\section{Analytic type I error control} \label{Analytic}
Given the max-type edge-count statistic, the next question is how large does the critical value need to be to constitute sufficient evidence against the null hypothesis of homogeneity (\ref{H0}). {{This is usually achieved by computing the $p$-value, which is defined as the probability of observing a more extreme or as extreme test statistic when the null hypothesis is ture. Here, the $p$-value is defined under the permutation null distribution of the max-type edge-count statisitc. As a relatively large value is the evidence for potential change-points,}} we are concerned with the tail probability of the test statistics under $H_0$:
\begin{eqnarray} \label{M}
\P\Big(\max_{n_0\leq t\leq n_1}M(t)>b\Big)
\end{eqnarray}

For a small $n$, the probability (\ref{M}) can be obtained directly by permutation. However, when $n$ is large, doing permutations could be very time-consuming. Hence, we derive analytic formulas to approximate the probability based on the asymptotic properties of the test statistic \eqref{eq:decomp}. We first work out the limiting distributions of $\{Z_w([nu]\footnote{For a scalar $x$, we use $[x]$ to denote the largest integer no greater than $x$.}):0<u<1\}$ and $\{Z_{\text{diff}}([nu]):0<u<1\}$ jointly.  On a directed graph, we use $e=(e_-,e_+)$ to denote an edge connected from $e_-$ to $e_+$.  Let $A_e=G_{e_-}\cup G_{e_+}$ be the subgraph in $G$ that connect to either node $e_-$ or node $e_+$, and $B_e=\cup_{e^*\in A_e}A_{e^*}$ be the subgraph in $G$ that connect to any edge in $A_e$. In the following, we write $a_n=O(b_n)$ when $a_n$ has the same order as $b_n$, and write $a_n=o(b_n)$ when $a_n$ has order smaller than $b_n$.

\begin{theorem} \label{theorem3}
For a directed $k$-NN graph, if $k=O(n^\beta)$, $\beta < 0.25$, $\sum_{e\in G}|A_e||B_e|=o(n^{1.5(\beta+1)})$, $\sum_{e\in G}|A_e|^2=o(n^{\beta+1.5})$, and $\sum_{i=1}^n |D_i|^2 - k^2 n = O(\sum_{i=1}^n |D_i|^2)$, as $n\rightarrow\infty$, $\{Z_w([nu]):0<u<1\}$ and $\{Z_{\text{diff}}([nu]):0<u<1\}$ converge to independent Gaussian processes in finite dimensional distributions.
\end{theorem}

The covariance functions of the limiting processes $\{Z_w([nu]):0<u<1\}$ and $\{Z_{\text{diff}}([nu]):0<u<1\}$ are provided in Supplement B.

\begin{figure}[h]
\center
\includegraphics[scale=0.27]{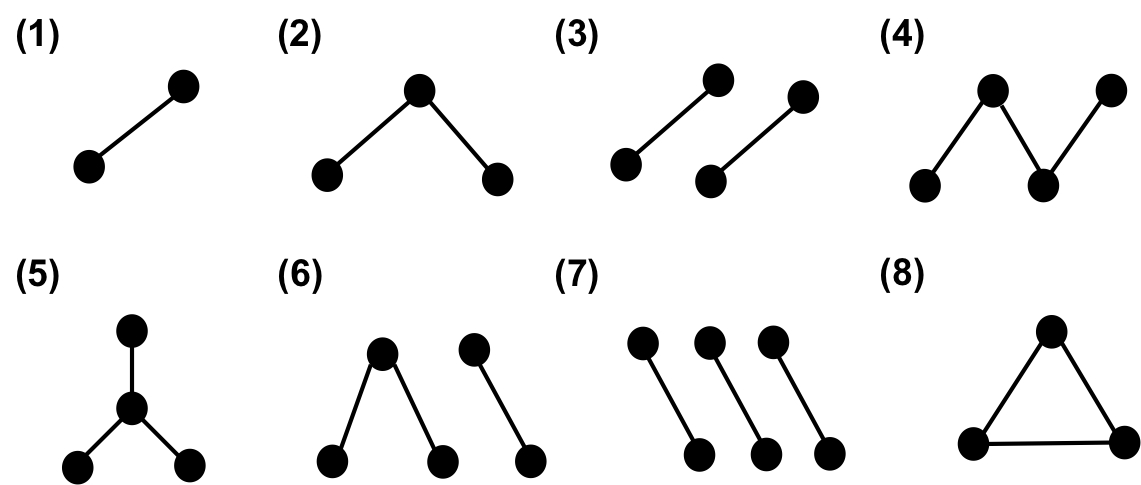}
\caption{Eight possible configurations of three edges randomly chosen with replacement from an undirected graph.}
\label{figure3}
\end{figure}

\begin{figure}[h]
\center
\includegraphics[scale=0.27]{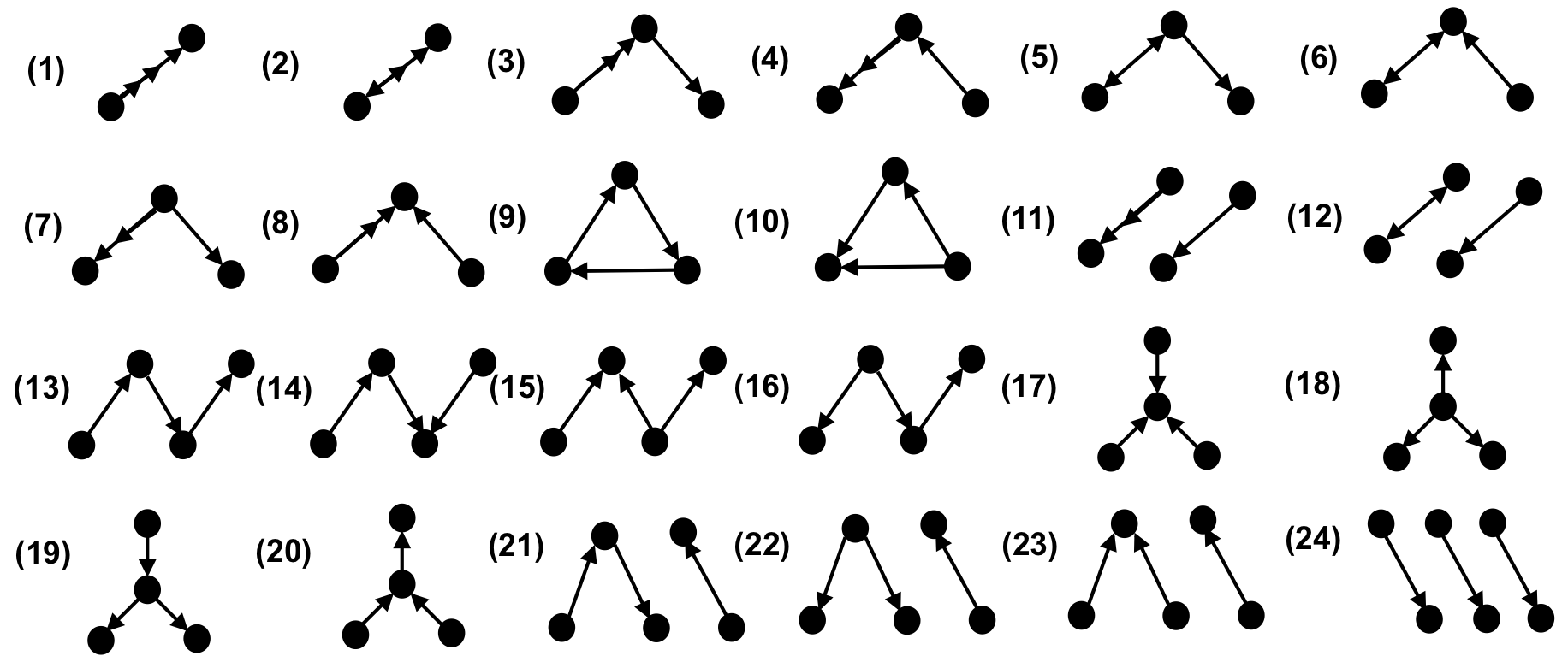}
\caption{Twenty-four possible configurations of three edges randomly chosen with replacement from a directed graph.}
\label{figure2}
\end{figure}

The complete proof for Theorem \ref{theorem3} is in Supplement A.3. {{The key idea of the proof is to decouple the dependency resulted from the permutation null distribution and the dependency caused by the graph.  More specifically, there are weak dependencies caused by the permutation as one observation appear at one time cannot appear at another time under permutation.  To solve this, we take a step back and work on the bootstrap null distribution in which the probability of an observation appearing at one time does not affect by whether it appears at other time(s) or not.   Thus, we could focus on dealing with the dependency caused by the graph, and the Stein's method is used to deal with the dependency.
The bootstrap null distribution is then connected to the permutation null distribution by conditioning.
}} Based on Theorem \ref{theorem3}, the probability (\ref{M}) can be approximated by
\begin{flalign} \label{eq:decomp}
& \P\Big(\max_{n_0\leq t\leq n_1}M(t)>b\Big) \\
& \approx 1-\P\Big(\max_{n_0\leq t\leq n_1}Z_w(t)<b\Big)\P\Big(\max_{n_0\leq t\leq n_1}|Z_{\text{diff}}(t)|<b\Big) \nonumber \\
& = 1-\left(1-\P\Big(\max_{n_0\leq t\leq n_1}Z_w(t)>b\Big)\right) \times \nonumber \\
& \hspace{20mm} \left(1-\P\Big(\max_{n_0\leq t\leq n_1}|Z_{\text{diff}}(t)|>b\Big)\right). \nonumber
\end{flalign}
The two probabilities in (\ref{eq:decomp}) can be computed similarly as in \cite{Chu2019}:
\begin{flalign}
& \hspace{-5mm} \P\Big(\max_{n_0\leq t\leq n_1}Z_w(t)>b\Big) \label{eq:Zw} \\
& \approx b\phi(b)\int_{n_0}^{n_1}S_w(t)C_w(t)\nu\big(\sqrt{2b^2C_w(t)}\big)dt \nonumber \\
& \hspace{-5mm} \P\Big(\max_{n_0\leq t\leq n_1}|Z_{\text{diff}}(t)|>b\Big) \label{eq:Zd} \\
& \approx 2b\phi(b)\int_{n_0}^{n_1}S_{\text{diff}}(t)C_{\text{diff}}(t)\nu\big(\sqrt{2b^2C_{\text{diff}}(t)}\big)dt \nonumber
\end{flalign}
where the function $\nu(\cdot)$ can be estimated numerically as $\nu(x)\approx\frac{(2/x)(\Phi(x/2)-0.5)}{(x/2)\Phi(x/2)+\phi(x/2)}$ with $\phi(\cdot)$ and $\Phi(\cdot)$ being the probability density function and cumulative distribution function of the standard normal distribution, respectively; $C_w(t)$, $C_{\text{diff}}(t)$ the partial derivative of the covariance function of the process; $S_w(t)$, $S_{\text{diff}}(t)$ the time-dependent skewness correction terms, i.e., for $j = w, \text{diff}$,
\begin{eqnarray*}
C_j(t) &=& \lim_{s\nearrow t}\frac{\partial \rho_j(s,t)}{\partial s}, \quad \rho_j(s,t)=\Cov(Z_j(s),Z_j(t)), \\
S_j(t) &=& \frac{\exp\big(\frac{1}{2}(b-\hat\theta_{b,j}(t))^2+\frac{1}{6}\gamma_j(t)\hat\theta^3_{b,j}(t))\big)}{\sqrt{1+\gamma_j(t)\hat\theta_{b,j}(t)}};
\end{eqnarray*}
where
\begin{eqnarray*}
\gamma_j(t)=\E\left(Z^3_j(t)\right), \quad \hat\theta_{b,j}(t)=(-1+\sqrt{1+2b\gamma_j(t)})/\gamma_j(t).
\end{eqnarray*}
Moreover, in the above expressions, $C_w(t)$ and $C_{\text{diff}}(t)$ can be derived and simplified to be (details in Supplement B)
\begin{eqnarray*}
C_w(t) = \frac{n(n-1)(2t^2/n-2t+1)}{2t(n-t)(t^2-nt+n-1)}, \quad
C_{\text{diff}}(t) = \frac{n}{2t(n-t)}.
\end{eqnarray*}
To compute $S_w(t)$ and $S_{\text{diff}}(t)$, we need the third moments of $Z_w(t)$ and $Z_{\text{diff}}(t)$, respectively. Comparing to that under undirected graphs, the computation here is much more complicated. For an undirected graph, there are 8 possible configurations (Fig. \ref{figure3}). However, for a directed graph, there are 24 possible configurations (Fig. \ref{figure2}). With brute force, it would take $O(|G|^3)$ time to compute the numbers of those configurations for a generic directed graph, which is very computationally expensive. To tackle this problem, we work out efficient formulas that can provide the results in $O(nk^2)$ time for directed $k$-NN graphs.

Let $G^{(m)}$ be the set of pairs of edges in $G$ having the $m$th configuration as shown in Fig. \ref{figure1}, $m=1,\ldots,7$.  Let $N^{(l)}$ be the number of occurrence for each of the configurations illustrated in Fig. \ref{figure2}, $l=1,\ldots,24$, then $\sum_{l=1}^{24}N^{(l)}=|G|^3$.  We can obtain $N^{(l)}$'s with effort:

\begin{flalign*}
& N^{(1)} = nk, \\
& N^{(2)} = 3c^{(2)}, \\
& N^{(3)} = 3c^{(3)}, \\
& N^{(4)} = 3c^{(6)}, \\
& N^{(5)} = 6c^{(2)}(k-1), \\
& N^{(6)} = 3\sum_{(i,j),(u,v)\in G^{(2)}}(|D_i|+|D_j|-2), \\
& N^{(7)} = 3c^{(5)}, \\
& N^{(8)} = 3c^{(3)}, \\
& N^{(9)} = 2\sum_{(i,j),(u,v)\in G^{(3)}}\mathbbm{1}_{\{(v,i)\in G\}}, \\
& N^{(10)} = 6\sum_{(i,j),(u,v)\in G^{(3)}}\mathbbm{1}_{\{(i,v)\in G\}}, \\
& N^{(11)} = 3c^{(7)}, \\
& N^{(12)} = 3c^{(2)}(nk-2)-(N^{(5)}+N^{(6)}), \\
& N^{(13)} = 6kc^{(3)}-(N^{(6)}+3N^{(9)}), \\
& N^{(14)} = 6\sum_{(i,j),(u,v)\in G^{(3)}}(|D_v|-1)-N^{(10)}, \\
& N^{(15)} = 6(k-1)c^{(6)}-N^{(10)}, \\
& N^{(16)} = 6kc^{(5)}-(N^{(5)}+N^{(10)}), \\
& N^{(17)} = 6\sum_{i=1}^n \binom{|D_i|}{3}, \\
& N^{(18)} = 6n\binom{k}{3}, \\
& N^{(19)} = 3\sum_{(i,j),(u,v)\in G^{(5)}}|D_i| - N^{(5)}, \\
& N^{(20)} = 3kc^{(6)}-N^{(6)}, \\
& N^{(21)} = 6c^{(3)}(nk-2) - \Big(N^{(5)}+N^{(6)}+N^{(10)}+N^{(14)} \\
& \hspace{15mm} +N^{(16)}+3N^{(9)}+2N^{(13)}+2N^{(19)}+2N^{(20)}\Big), \\
& N^{(22)} = 3c^{(5)}(nk-2) \\
& \hspace{5mm} - \left(N^{(5)}+N^{(10)}+N^{(15)}+N^{(16)}+N^{(19)}+3N^{(18)}\right), \\
& N^{(23)} = 3c^{(6)}(nk-2) \\
& \hspace{5mm} - \left(N^{(6)}+N^{(10)}+N^{(14)}+N^{(15)}+N^{(20)}+3N^{(17)}\right), \\
& N^{(24)} = (nk)^3 - \sum_{l=1}^{23}N^{(l)}.
\end{flalign*}

In the above formulas, it takes at most $O(nk^2)$ time to compute $c^{(2)},c^{(3)},c^{(5)}$, and $c^{(6)}$, and there are at most $nk^2$ elements in the sets $G^{(2)},G^{(3)}$, and $G^{(5)}$, so the numbers of the 24 configurations can be calculated within $O(nk^2)$ time. Indeed, as the rest of the computation is relatively straightforward (see Supplement C), the whole analytic $p$-value approximation procedure for a directed $k$-NN graph can also be done within $O(nk^2)$.

Now, let's check the performance of (\ref{eq:decomp}) through simulation studies.

\begin{table}[h]
\center
\caption{Critical values for the statistic $\displaystyle{\max_{n_0\leq t\leq n_1}M(t)}$ based on 3-NN's graph at $\alpha=0.05$.}
\label{type_I}

\begin{tabular}{p{3mm}p{4mm}|llllllll}
\hline
 & & \multicolumn{8}{c}{Critical Values} \\ [0.1cm]
 & & \multicolumn{2}{c}{$n_0=100$} & \multicolumn{2}{c}{$n_0=75$} & \multicolumn{2}{c}{$n_0=50$} & \multicolumn{2}{c}{$n_0=25$} \\ [0.1cm]
 & $d$ & Ana & Per & Ana & Per & Ana & Per & Ana & Per \\ [0.2cm] \hline
\multirow{3}{*}{\shortstack[c]{(C1) \\ \strut}}
 & 10 & 3.26 & 3.26 & 3.31 & 3.35 & 3.39 & 3.43 & 3.52 & 3.60 \\
 & 100 & 3.29 & 3.29 & 3.36 & 3.40 & 3.45 & 3.52 & 3.62 & 3.78 \\
 & 1,000 & 3.31 & 3.31 & 3.40 & 3.42 & 3.51 & 3.60 & 3.70 & 3.94 \\ [0.2cm]
\multirow{3}{*}{\shortstack[c]{(C2) \\ \strut}}
 & 10 & 3.26 & 3.26 & 3.32 & 3.34 & 3.39 & 3.44 & 3.51 & 3.60 \\
 & 100 & 3.30 & 3.30 & 3.35 & 3.39 & 3.44 & 3.51 & 3.60 & 3.79 \\
 & 1,000 & 3.30 & 3.30 & 3.46 & 3.54 & 3.59 & 3.75 & 3.80 & 4.27 \\ [0.2cm]
\multirow{3}{*}{\shortstack[c]{(C3) \\ \strut}}
 & 10 & 3.27 & 3.28 & 3.32 & 3.33 & 3.40 & 3.41 & 3.52 & 3.58 \\
 & 100 & 3.28 & 3.28 & 3.35 & 3.37 & 3.43 & 3.48 & 3.58 & 3.70 \\
 & 1,000 & 3.36 & 3.41 & 3.45 & 3.58 & 3.58 & 3.81 & 3.72 & 4.07 \\ \hline
\end{tabular}
\end{table}

Table \ref{type_I} shows the performance of the asymptotic $p$-value approximation of the max-type edge-count statistic (\ref{eq:decomp}) under different settings. We examine three different distributions (multivariate Gaussian (C1), multivariate $t_5$ (C2), and multivariate log-normal (C3) distributions) with different data dimensions ($d=10,100,1000$). In Table \ref{type_I}, column ``Per" is the critical value obtained from doing $10,000$ permutations.  This can be deemed as close to the true critical value. Column ``Ana" presents the analytical critical values given by plugging \eqref{eq:decomp} with \eqref{eq:Zw} and \eqref{eq:Zd}. Here, the length of the sequence is $n=1000$, and we present the results in four different choices of $n_0$ with $n_1=n-n_0$. When $n_0=100$ or $75$, the analytical approximation works quite well across all distributions and dimensions. As $n_0$ decreases ($n_0=50$ or $25$), the analytical critical values become less precise. This is expected as the asymptotic distribution needs both groups to have $O(n)$ observations. When $n_0$ is small, one group could give a smaller order of observations than the other group. On the other hand, the accuracy of the analytical critical values is less dependent on the distribution of the data.

\section{Numerical results} \label{Section.4}

\subsection{Simulation setting and notations}

We compare our proposed test with three state-of-the-art methods: graph-based method with the max-type edge-count statistic on $5$-MST (the recommended setting in \cite{Chu2019}, denoted by ``5-MST" in the following), the distance-based method (ecp) \cite{Matteson2014}, and the kernel-based method \cite{H2019}. For our method, we use the $kd$-tree algorithm \cite{FNN} to approximate the directed $5$-NN graph (d-a$5$NN). {{In the following, we focus on the Euclidean distance.  There are implementations for other Minkowski distances in the \texttt{ANN} library (\url{http://www.cs.umd.edu/~mount/ANN/})}}. We use the directed approximate $5$-NN graph as it contains a similar number of edges to the $5$-MST to make the comparison with the existing graph-based method fair. The proposed method is denoted by ``New" in the following.

\subsection{Computational efficiency} \label{Time}

First, we compare the computational cost of these methods through $10$ simulation runs. In each simulation, the observations are generated i.i.d. from a multivariate Gaussian distribution with dimension $d=500$. The results are presented in Table \ref{time_comparison}. Among all the four methods, our proposed method is the fastest, whereas the kernel method is the slowest to run. For the graph-based methods, in particular, our proposed method on d-a$5$NN is more than $5$ times faster than the method in \cite{Chu2019} on $5$-MST for $n=2,000$ or above.

\begin{table}[h]
\center \small
\caption{Runtime comparison: Average time cost in seconds (standard deviation) from {{$100$}} simulation runs for each choice of $n$ {{($10$ runs for the cells having average runtime greater than 1k seconds)}}. The environment where the experiments are conduected: CPU: Intel(R) Xeon(R) CPU E5-2690 0 @ 2.90GHz / RAM: DDR3 @ 1600MHz / OS: Scientific Linux 6.10 / 2.6.32 Linux.}
\label{time_comparison}
\begin{tabular}{l|llll}
\hline
 \multicolumn{1}{c|}{$n$} & New & 5-MST & ecp & kernel \\ [0.25cm] \hline
 $1,000$ & 0.8 ($0.01$) & 5.2 ($0.5$) & 13 ($1.8$) & 683 ($17$) \\ [0.2cm]
 $2,000$ & 2.7 ($0.01$) & 21 ($3.2$) & 52 ($6.7$) & 10,224 \\ [0.2cm] 
 $5,000$ & 17 ($0.1$) & 157 ($8.3$) & 482 ($87$) & $>$10,000 \\ [0.2cm]
 $10,000$ & 76 ($1.2$) & 689 ($27$) & 2,073 ($387$) & - \\ [0.2cm]
 $20,000$ & 321 ($3.7$) & 2,193 ($189$) & 6,528 ($1,276$) & - \\ [0.2cm]
 $30,000$ & 726 ($5.9$) & 4,757 ($106$) & $>$10,000 & - \\ [0.1cm] \hline
\end{tabular}
\end{table}

\subsection{Empirical size} \label{Size}

Here, we check the empirical size of these methods under three significance levels ($\alpha=0.10,0.05$, and $0.01$). Observations are generated i.i.d. from a $d$-dimensional multivariate Gaussian distribution with no change-point. The results are summerized in Table \ref{empirical_size}, {{where the $p$-value of the graph-based methods is obtained through their corresponding analytical formulas, and those for the ecp are based on 999 random permutations.}} We only report those for $d=25$ here, as the results are very similar for other choices of $d$ (see Supplement E). Both the graph-based methods and ecp could control the type I error well. However, for the kernel method, there is no direct mean to control type I error. We use the \texttt{`kcpa'} function in the R package \texttt{`ecp'}. In this function, the empirical size is supervised by a tuning parameter $C$. The larger the $C$ is, the less likely the null is rejected. Unfortunately, it is not straightforward to link the tuning parameter $C$ with the empirical size. As illustrated in Table \ref{empirical_kernel}, the relation between $C$ and the empirical size depends heavily on the dimension of the observations.

\begin{table}[h]
\center \small
\caption{Fractions of simulation runs (out of 10,000 simulations) that the null hypothesis is rejected when there is no change-point in the sequence ($n=1,000$). Graph-based methods and ecp at level $\alpha$.}
\label{empirical_size}

\begin{adjustbox}{scale=1.2}
\begin{tabular}{l|ccc}
\hline
 Method & $\alpha=0.10$ & $\alpha=0.05$ & $\alpha=0.01$ \\ [0.2cm] \hline
 New & $0.100$ & $0.051$ & $0.011$ \\
 5-MST & $0.096$ & $0.051$ & $0.012$ \\
 ecp & $0.098$ & $0.050$ & $0.011$ \\ \hline
\end{tabular}
\end{adjustbox}

\end{table}

\begin{table}[h]
\center \small
\caption{Fractions of simulation runs (out of 10,000 simulations) that the null hypothesis is rejected when there is no change-point in the sequence ($n=1,000$). Kernel method with tuning parameter $C$ under three different dimensions.}
\label{empirical_kernel}

\begin{adjustbox}{scale=1.2}
\begin{tabular}{c|ccc}
\hline
 kernel method & $C=4.8$ & $C=5.2$ & $C=5.6$ \\ [0.2cm] \hline
 $d=25$ & $0.943$ & $0.821$ & $0.631$ \\ 
 $d=30$ & $0.477$ & $0.265$ & $0.134$ \\ 
 $d=35$ & $0.094$ & $0.036$ & $0.009$ \\ \hline 
\end{tabular}
\end{adjustbox}

\end{table}

{{
\subsection{Type II error analysis} \label{Type2}
To get an idea of the performance of the proposed method, we compare the probability of making the type II error, which is the event that the null hypothesis is not rejected when it is false, for the three methods can could control the type I error under some common parametric families. 
In particular, we consider six scenarios with each coordinate randomly generated from (1) Chi-square distributions with a change in the degree of freedom, (2) Weibull distribution with a change in the scale parameter while the shape parameter is fixed, (3) \& (4) Gamma distributions with a change in one of the two parameters, respectively, while the other parameter is fixed, and (5) \& (6) Beta distributions with a change in one of the two parameters, respectively, while the other parameter is fixed.  In each simulation run, the length of the sequence is $n=1,000$, and the change-point is at a quarter of the sequence $\tau=250$. 

\begin{itemize}
\item Setting 1 (Chi-square distribution): $F_0=\chi^2_{\nu_0}$; $F_1=\chi^2_{\nu_1}$.

\item Setting 2 (Weibull distribution): $F_0=\text{Weibull}(\lambda_0,k_0)$; $F_1=\text{Weibull}(\lambda_1,k_0)$. Shape parameter fixed at $k_0=1$.

\item Setting 3 (Gamma distribution-a): $F_0=\text{Gamma}(\alpha_0,\beta_0)$; $F_1=\text{Gamma}(\alpha_1,\beta_0)$. Scale parameter fixed at $\beta_0=1$.

\item Setting 4 (Gamma distribution-b): $F_0=\text{Gamma}(\alpha_0,\beta_0)$; $F_1=\text{Gamma}(\alpha_0,\beta_1)$. Shape parameter fixed at $\alpha_0=1$.

\item Setting 5 (Beta distribution-a): $F_0=\text{Beta}(\alpha_0,\beta_0)$; $F_1=\text{Gamma}(\alpha_1,\beta_0)$. Second parameter fixed at $\beta_0=0.5$.

\item Setting 6 (Beta distribution-b): $F_0=\text{Beta}(\alpha_0,\beta_0)$; $F_1=\text{Gamma}(\alpha_0,\beta_1)$. First parameter fixed at $\alpha_0=0.5$.
\end{itemize}

\begin{table}[h]
\center
\caption{{{Type II error: Numbers of times (out of 100) the null hypothesis is not rejected under $\alpha=0.05$ for various data dimensions and sizes of change.}}}
\label{type2error}

\begin{adjustbox}{scale=1.2}
\begin{tabular}{l|ccccc}
\multicolumn{6}{c}{\B S1: Chi-square (d.f. $\nu$ change, $\nu_0=3$)} \\ [0.5mm] \hline
 \multicolumn{1}{c|}{$d$} & 25 & 100 & 500 & 1000 & 2000 \\
 \multicolumn{1}{c|}{$\nu_1$} & 3.27 & 3.20 & 3.15 & 3.12 & 3.09 \\ \hline
 New & \B.16 & \B.32 & \B.13 & .11 & .13 \\
 5-MST & .19 & .37 & \B.13 & \B.10 & \B.11 \\
 ecp & .95 & .95 & .95 & .95 & .96 \\ \hline
\end{tabular}
\end{adjustbox}

\vspace{0.3cm}

\begin{adjustbox}{scale=1.2}
\begin{tabular}{l|ccccc}
\multicolumn{6}{c}{\B S2: Weibull (scale $\lambda$ change, $\lambda_0=1,k_0=1$)} \\ [0.5mm] \hline
 \multicolumn{1}{c|}{$d$} & 25 & 100 & 500 & 1000 & 2000 \\
 \multicolumn{1}{c|}{$\lambda_1$} & 1.8 & 2.4 & 3.2 & 4.8 & 6.2 \\ \hline
 New & \B.19 & \B.17 & \B.11 & \B.18 & \B.23 \\
 5-MST & .24 & .19 & .13 & .19 & \B.23 \\
 ecp & .93 & .97 & .94 & .93 & .97 \\ \hline
\end{tabular}
\end{adjustbox}

\vspace{0.3cm}

\begin{adjustbox}{scale=1.2}
\begin{tabular}{l|ccccc}
\multicolumn{6}{c}{\B S3: Gamma (shape change, $\alpha_0=1,\beta_0=1$)} \\ [0.5mm] \hline
 \multicolumn{1}{c|}{$d$} & 25 & 100 & 500 & 1000 & 2000 \\
 \multicolumn{1}{c|}{$\alpha_1$} &  1.09 & 1.08 & 1.05 & 1.04 & 1.03 \\ \hline
 New & \B.12 & \B.15 & .34 & .28 & .33 \\
 5-MST & .19 & .18 & \B.29 & \B.22 & \B.28 \\
 ecp & .92 & .91 & .89 & .95 & .91 \\ \hline
\end{tabular}
\end{adjustbox}

\vspace{0.3cm}

\begin{adjustbox}{scale=1.2}
\begin{tabular}{l|ccccc}
\multicolumn{6}{c}{\B S4: Gamma (scale change, $\alpha_0=1,\beta_0=1$)} \\ [0.5mm] \hline
 \multicolumn{1}{c|}{$d$} & 25 & 100 & 500 & 1000 & 2000 \\
 \multicolumn{1}{c|}{$\beta_1$} & 1.050 & 1.040 & 1.030 & 1.025 & 1.020 \\ \hline
 New & \B.32 & \B.29 & .25 & .11 & .12 \\
 5-MST & .36 & .36 & \B.24 & \B.08 & \B.07 \\
 ecp & .95 & .93 & .92 & .90 & .89 \\ \hline
\end{tabular}
\end{adjustbox}

\vspace{0.3cm}

\begin{adjustbox}{scale=1.2}
\begin{tabular}{l|ccccc}
\multicolumn{6}{c}{\B S5: Beta (shape 1 change, $\alpha_0=0.5,\beta_0=0.5$)} \\ [0.5mm] \hline
 \multicolumn{1}{c|}{$d$} & 25 & 100 & 500 & 1000 & 2000 \\
 \multicolumn{1}{c|}{$\alpha_1$} & 0.590 & 0.550 & 0.530 & 0.520 & 0.512 \\ \hline
 New & .23 & \B.19 & \B.05 & \B.03 & \B.19 \\
 5-MST & \B.21 & .23 & .06 & .05 & .25 \\
 ecp & .97 & .96 & .94 & .96 & .95 \\ \hline
\end{tabular}
\end{adjustbox}

\vspace{0.3cm}

\begin{adjustbox}{scale=1.2}
\begin{tabular}{l|ccccc}
\multicolumn{6}{c}{\B S6: Beta (shape 2 change, $\alpha_0=0.5,\beta_0=0.5$)} \\ [0.5mm] \hline
 \multicolumn{1}{c|}{$d$} & 25 & 100 & 500 & 1000 & 2000 \\
 \multicolumn{1}{c|}{$\beta_1$} & 0.590 & 0.550 & 0.530 & 0.520 & 0.512 \\ \hline
 New & .25 & \B.14 & \B.04 & \B.05 & \B.16 \\
 5-MST & \B.24 & .18 & .06 & .09 & .24 \\
 ecp & .98 & .93 & .96 & .92 & .91 \\ \hline
\end{tabular}
\end{adjustbox}

\end{table}

The results are shown in Table \ref{type2error}.  For each dimension, the alternatives are chosen so that the type II error is not too small to be comparable.  We see that the type II error of the new test in on the small end for data from different distribution families.
}}

\subsection{Power comparison} \label{Power}

 Here, we compare the power of the proposed method to the other two methods. {{Power is the probability of rejecting the null hypothesis when it is false, i.e., the probability of not making the type II error. }} 
 We consider six different scenarios. 
 {{
They are chosen to cover a variety of change types. Scenarios 1-4 emphasize on the Gaussian distribution and cover changes in mean and variance, as well as different parts of the covariance matrix.  Scenarios 5 and 6 cover asymmetric distributions and fat-tailed distributions.
}} In the following, $a$ and $b$ are constants. $N_d$ denotes a $d$-dimensional multivariate Gaussian distribution, $\mathbf{0}_d$ and $\mathbf{1}_d$ denote length-$d$ vectors of all zeros and one's, respectively, $\mathbbm{I}_d$ denotes a $d\times d$ identity matrix, and $\bm{\Sig}$ denotes the covariance matrix with $\bm{\Sig}_{ij}=0.6^{|i-j|}$, where $\bm{\Sig}_{ij}$ is the element of the $i$th row and the $j$th column of $\bm{\Sig}$. The $L_2$ norm of the mean vector in $F_1$ is given by $||\Delta||_2$ in Table \ref{power_comparison}.  In each simulation run, the length of the sequence is $n=1,000$, and the change-point is at a quarter of the sequence $\tau=250$. 

\begin{table}[h]
\center
\caption{Power comparison: Numbers of times (out of 100) the null hypothesis is rejected under $\alpha=0.05$ for various data dimensions and sizes of change.}
\label{power_comparison}

\begin{adjustbox}{scale=1.2}
\begin{tabular}{l|ccccc}
\multicolumn{6}{c}{\B S1: MG (Mean and Variance)} \\ [0.5mm] \hline
 \multicolumn{1}{c|}{$d$} & 25 & 100 & 500 & 1000 & 2000 \\
 \multicolumn{1}{c|}{$||\Delta||_2$} & 0.10 & 0.20 & 0.45 & 0.63 & 0.89 \\
 \multicolumn{1}{c|}{$b$} & 1.10 & 1.06 & 1.03 & 1.02 & 1.02 \\ \hline
 New & \B75 & \B76 & \B65 & \B58 & \B83 \\
 5-MST & 71 & 66 & 60 & 56 & \B83 \\
 ecp & 4 & 8 & 10 & 13 & 15 \\ \hline
\end{tabular}
\end{adjustbox}

\vspace{0.3cm}

\begin{adjustbox}{scale=1.2}
\begin{tabular}{l|ccccc}
\multicolumn{6}{c}{\B S2: MG (5-coordinate)} \\ [0.5mm] \hline
 \multicolumn{1}{c|}{$d$} & 25 & 100 & 500 & 1000 & 2000 \\
 \multicolumn{1}{c|}{$||\Delta||_2$} & 0.20 & 0.40 & 0.67 & 0.63 & 0.89 \\
 \multicolumn{1}{c|}{$b$} & 1.8 & 2.4 & 3.2 & 4.8 & 6.2 \\ \hline
 New & \B94 & \B84 & \B63 & 64 & 69 \\
 5-MST & 92 & 83 & 60 & \B65 & \B71 \\
 ecp & 17 & 36 & 33 & 22 & 34 \\ \hline
\end{tabular}
\end{adjustbox}

\vspace{0.3cm}

\begin{adjustbox}{scale=1.2}
\begin{tabular}{l|ccccc}
\multicolumn{6}{c}{\B S3: MG (Diagonal)} \\ [0.5mm] \hline
 \multicolumn{1}{c|}{$d$} & 25 & 100 & 500 & 1000 & 2000 \\
 \multicolumn{1}{c|}{$b$} &  1.10 & 1.06 & 1.03 & 1.02 & 1.02 \\ \hline
 New & \B69 & \B78 & 87 & \B90 & \B99 \\
 5-MST & 60 & 77 & \B88 & 88 & \B99 \\
 ecp & 3 & 7 & 4 & 2 & 4 \\ \hline
\end{tabular}
\end{adjustbox}

\vspace{0.3cm}

\begin{adjustbox}{scale=1.2}
\begin{tabular}{l|ccccc}
\multicolumn{6}{c}{\B S4: MG (Off-diagonal)} \\ [0.5mm] \hline
 \multicolumn{1}{c|}{$d$} & 25 & 100 & 500 & 1000 & 2000 \\
 \multicolumn{1}{c|}{$\rho$} & 0.53 & 0.50 & 0.48 & 0.47 & 0.46 \\ \hline
 New & \B74 & \B88 & \B89 & \B88 & \B89 \\
 5-MST & 68 & 83 & 87 & 87 & 88 \\
 ecp & 4 & 9 & 8 & 4 & 3 \\ \hline
\end{tabular}
\end{adjustbox}

\vspace{0.3cm}

\begin{adjustbox}{scale=1.2}
\begin{tabular}{l|ccccc}
\multicolumn{6}{c}{\B S5: Chi-square distribution} \\ [0.5mm] \hline
 \multicolumn{1}{c|}{$d$} & 25 & 100 & 500 & 1000 & 2000 \\
 \multicolumn{1}{c|}{$||\Delta||_2$} & 0.05 & 0.10 & 0.22 & 0.32 & 0.45 \\
 \multicolumn{1}{c|}{$b$} & 1.10 & 1.08 & 1.05 & 1.04 & 1.03 \\ \hline
 New & \B71 & \B81 & \B85 & \B85 & \B89 \\
 5-MST & 66 & 80 & \B85 & 83 & \B89 \\
 ecp & 4 & 10 & 3 & 7 & 2 \\ \hline
\end{tabular}
\end{adjustbox}

\vspace{0.3cm}

\begin{adjustbox}{scale=1.2}
\begin{tabular}{l|ccccc}
\multicolumn{6}{c}{\B S6: $t$-distribution} \\ [0.5mm] \hline
 \multicolumn{1}{c|}{$d$} & 25 & 100 & 500 & 1000 & 2000 \\
 \multicolumn{1}{c|}{$||\Delta||_2$} & 0.20 & 0.40 & 0.67 & 0.63 & 0.89 \\
 \multicolumn{1}{c|}{$b$} & 1.14 & 1.08 & 1.05 & 1.04 & 1.03 \\ \hline
 New & \B85 & \B75 & \B73 & 86 & 85 \\
 5-MST & 81 & 73 & 70 & \B87 & \B87 \\
 ecp & 8 & 15 & 18 & 11 & 11 \\ \hline
\end{tabular}
\end{adjustbox}

\end{table}

\begin{itemize}
\item Scenario 1 (MG: mean and variance): $F_0=N_d(\mathbf{0}_d,\bm{\Sig})$; $F_1=N_d(a\times \mathbf{1}_d,b\bm{\Sig})$.

\item Scenario 2 (MG: 5-coordinate): $F_0=N_d(\mathbf{0}_d,\mathbbm{I}_d)$; $F_1=N_d((a\times \mathbf{1}_5,\mathbf{0}_{d-5})^T,\text{diag}((b\times \mathbf{1}_5,\mathbf{1}_{d-5})^T))$, where $\text{diag}(u)$ is a diagonal matrix with its diagonal vector $u$.

\item Scenario 3 (MG: Diagonal): $F_0=N_d(\mathbf{0}_d,\mathbbm{I}_d)$; $F_1=N_d(\mathbf{0}_d,b\mathbbm{I}_d)$.

\item Scenario 4 (MG: Off-diagonal): $F_0=N_d(\mathbf{0}_d,\bm{\Sig})$; $F_1=N_d(\mathbf{0}_d,\bm{\Sig'})$, where $\bm{\Sig'}_{ij}=\rho^{|i-j|}$.

\item Scenario 5 (Chi-square distribution): $F_0=\bm{\Sig}^{\frac{1}{2}}\mathbf{u}^{\chi^2_{3,c}}$; $F_1=(b\bm{\Sig})^{\frac{1}{2}}\mathbf{u}^{\chi^2_{3,c}}+a\times \mathbf{1}_d$. Here, $\mathbf{u}^{\chi^2_{3,c}}$ is a length-$d$ vector with each component i.i.d. from the centered $\chi^2_3$ distribution.

\item Scenario 6 ($t$-distribution): $F_0=\bm{\Sig}^{\frac{1}{2}}\mathbf{u}^{t_5}$; $F_1=(b\bm{\Sig})^{\frac{1}{2}}\mathbf{u}^{t_5}+a\times \mathbf{1}_d$. Here, $\mathbf{u}^{t_5}$ is a length-$d$ vector with each component i.i.d. from the $t_5$-distribution.
\end{itemize}

From Table \ref{power_comparison}, we can see that our proposed test has good power under a wide range of alternatives. It is the best or on par with the best in these simulation studies. {{In sharp contrast, the ecp method suffers from the curse of dimensionality, and could have low power when the change contain sources other than the mean shift.}}

{{

\subsection{Types of changes the new method can detect} \label{Various}
Here, we check the types of changes the proposed method can detect.  We investigate five types of changes: mean, variance, covariance, skewness and kurtosis. All experiments are conducted under the setting with $n=1,000$, $\tau = 250$ and $d=1,000$. 
 Below describes in details the five scenrios:

\begin{itemize}
\item Change in mean: Before the change, all coordinates are from independent standard Gaussian distributions. After the change, the $L_2$-norm of the mean vector is specified by the x-axis in Fig. \ref{sensitivity_mean}. We study the power for changes in all coordinates ($dc=1,000$), one coordinate ($dc=1$), and some subsets of the coordinates ($dc=200,50,10$). We can see from Fig. \ref{sensitivity_mean} that our test has good power to mean change regardless of the number of coordinates that has a mean shift.

\begin{figure}[h]
\centering
\includegraphics[scale=0.50]{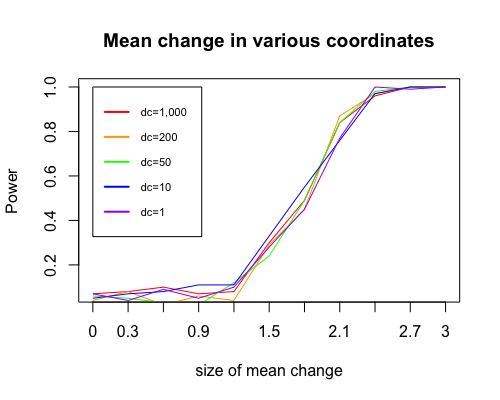}
\caption{Fraction of times (out of 100) that a change-point is detected at a given size of mean change for the changes in various coordinates (dc).}
\label{sensitivity_mean}
\end{figure}

\item Change in variance: Before the change, all coordinates are from independent standard Gaussian distributions. After the change, the determinant of the variance-covariance matrix becomes $a^{1000}$, where $a$ is specified by the x-axis in Fig. \ref{sensitivity_var}. We study the power for changes in all coordinates ($dc=1,000$), one coordinate ($dc=1$), and some subsets of the coordinates ($dc=200,50,10$). We can see from Fig. \ref{sensitivity_var} that our test is generally sensitive to variance change in all cases, and the power is higher when the change comes in fewer coordinates.

\begin{figure}[h]
\centering
\includegraphics[scale=0.50]{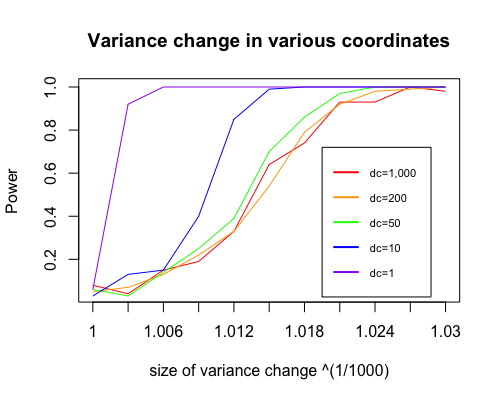}
\caption{Fraction of times (out of 100) that a change-point is detected at a given size of variance change for the changes in various coordinates (dc).}
\label{sensitivity_var}
\end{figure}

\item Change in covariance: Before the change, the observations are from multivariate Gaussian distribution with mean zero and variance-covariance matrix $\bm{\Sig}_{ij}=0.6^{|i-j|}$, denoted as $\rho_0=0.6$. After the change, the variance-covariance matrix becomes $\bm{\Sig}_{ij}=\rho_1^{|i-j|}$ and new correlation coefficent is defined as $\rho_1=0.6-\Delta\rho$. The ten values of the $\Delta\rho$'s used in the experiment are ($0.02,0.04,\ldots,0.20$), corresponding to the x-axis ($1,2,\ldots,10$) in Fig. \ref{sensitivity_rest}. The result shows that our new method is also very sensitive to changes in the covariance structure.

\item Change in skewness: Before the change the observations in each coordinate are from independent Gaussian distributions with mean $\nu$ and standard deviation $\sqrt{2\nu}$. After the change, observations in each coordinate are from independent Chi-square distributions with degree of freedom $\nu$. The skewness of a $\chi^2_\nu$ distribution is computed as $\sqrt{8/\nu}$. The ten values of the $\nu$'s used in the experiment are chosen so that the skewness of each variable are ($0.2,0.4,0.6,\ldots,2.0$), corresponding to the x-axis ($1,2,\ldots,10$) in Fig. \ref{sensitivity_rest}.  This setting does not change mean and variance while changing the skewness.

\item Change in excess kurtosis: Before the change the observations in each coordinate are from independent standard Gaussian distributions. After the change, observations in each coordinate are from independent $t$ distributions with degree of freedom $\nu$. The excess kurtoosis of a $t_\nu$ distribution is computed as $\frac{6}{\nu-4}$. The ten values of the $\nu$'s used in the experiment are chosen so that the excess kurtosis of each variable are ($0.01,0.02,0.03,\ldots,0.10$), corresponding to the x-axis ($1,2,\ldots,10$) in Fig. \ref{sensitivity_rest}. The result shows that our method can also detect changes in excess kurtosis.

\begin{figure}[h]
\centering
\includegraphics[scale=0.50]{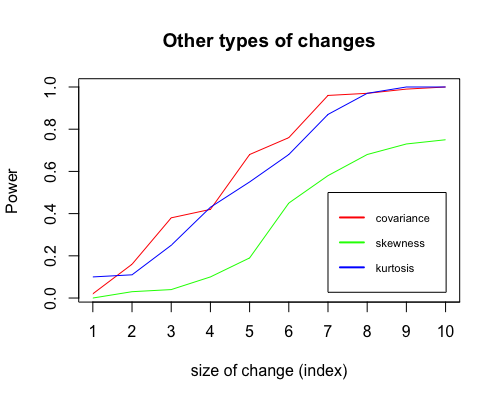}
\caption{Fraction of times (out of 100) that a change-point is detected at a given size of changes in covariance, skewness and excess kurtosis. The sizes of changes increase as the index in the x-axis grows, but the sizes among the three scenarios at each index are not comparable. We plot them on the same figure to save space.}
\label{sensitivity_rest}
\end{figure}
\end{itemize}

}}


\section{Real data applications} \label{application}

\subsection{fMRI data} \label{Section5.1}

This fMRI dataset was recorded when the subjects were watching certain pieces of the movie ``The Grand Budapest Hotel" by Wes Anderson. It is publicly available at: \texttt{\url{https://openneuro.org/datasets/ds003017/versions/1.0.2}}. There are in total 25 subjects involved in this experiment, each of them watching 5 pieces of the movie \cite{Budapest}. Here, we  {{randomly select two such sequences with subject ID SID-000005 and SID-000024}} for illustration. 

\begin{figure}[h]
\center
\includegraphics[scale=0.22]{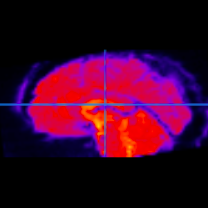}
\includegraphics[scale=0.22]{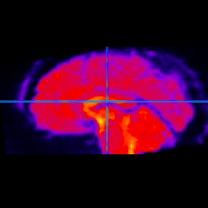}
\includegraphics[scale=0.22]{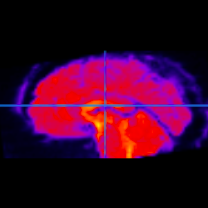}
\includegraphics[scale=0.22]{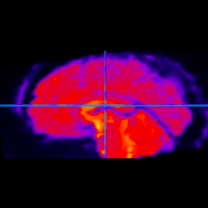}
\includegraphics[scale=0.22]{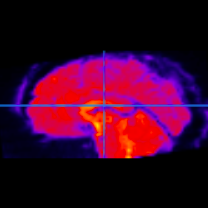}
$t=150$\hspace{6.1mm}$t=250$\hspace{6.1mm}$t=350$\hspace{6.1mm}$t=450$\hspace{6.1mm}$t=550$
\caption{The snapshots of the fMRI images for subject SID-000005 at different timestamps.}
\label{snapshots1}
\end{figure}

This piece of the movie is about 10 minutes long. The total length of the sequence is $n=598$ with one time unit as 1 second. Each observation is a 3-dimensional fMRI image with size $96\times 96\times 48$. To get an idea of how the fMRI data look like, Fig. \ref{snapshots1} shows the profiles of five observations at $t=150,250,350,450$, and $550$, from one certain perspective. There are three different perspectives available, and the other two can be found in Supplement F. We rearrange each observation into a length-$d$ vector ($d=96\times 96\times 48=442,368$).

\begin{figure}[h]
\center
\includegraphics[scale=0.18]{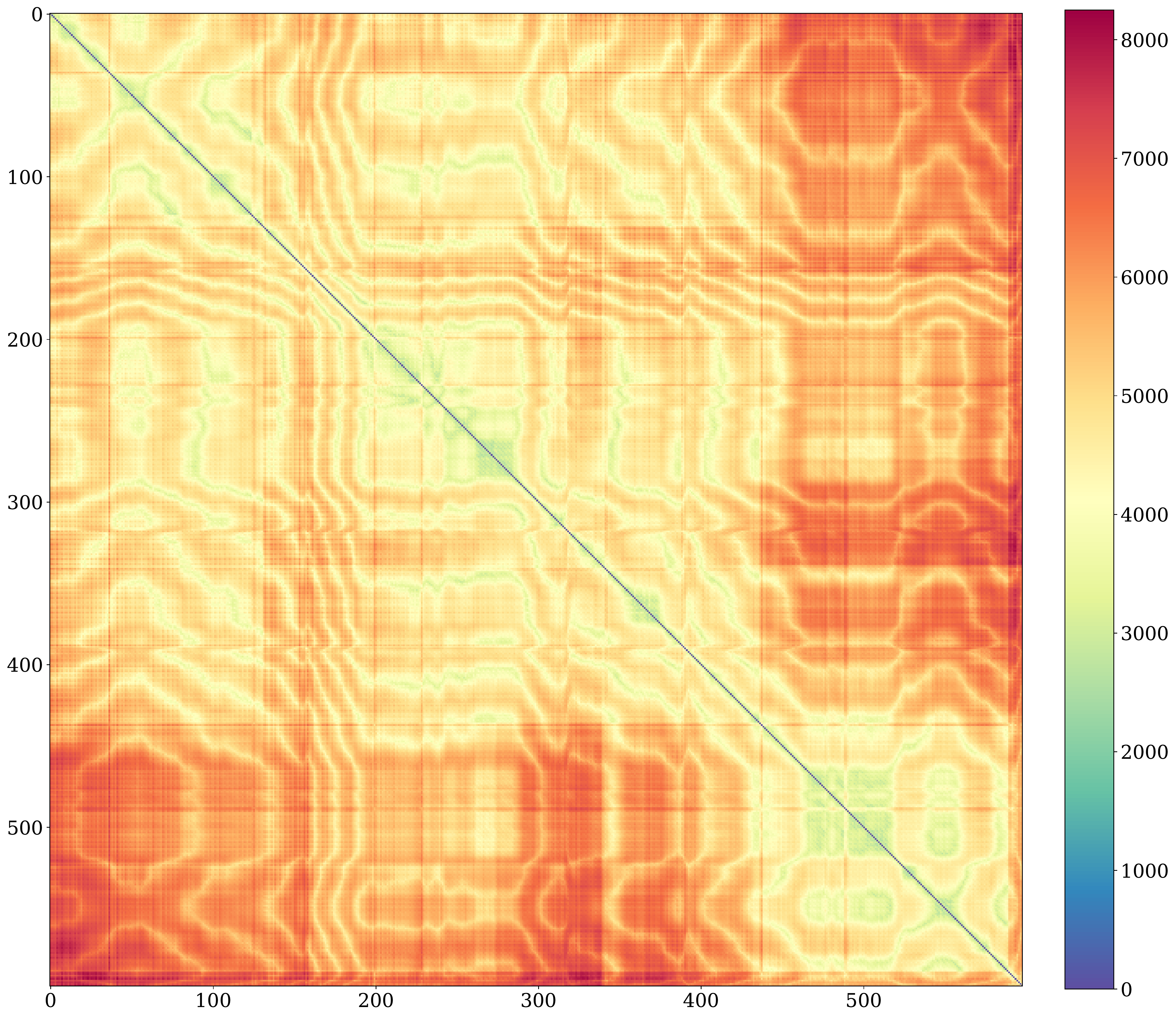}
\caption{Heatmap of pairwise distances of the observations in the sequence (SID-000005).}
\label{Heatmap_1}
\end{figure}

\begin{figure}[!]
\center
\includegraphics[scale=0.18]{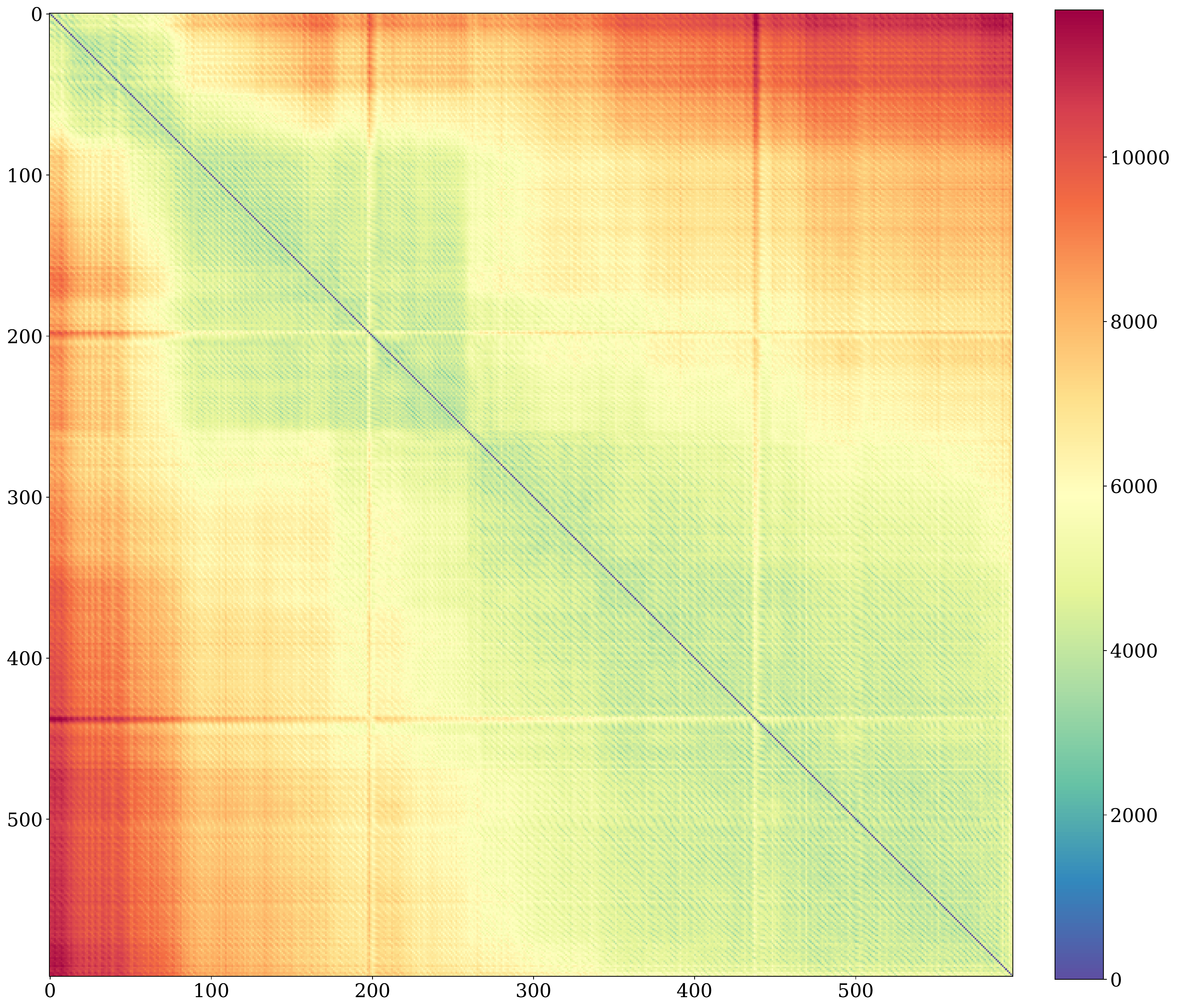}
\caption{Heatmap of pairwise distances of the observations in the sequence (SID-000024).}
\label{Heatmap_8}
\end{figure}

In the first sequence (SID-000005), the signal is strong that all three methods find the change-point at around $435$ (see Table \ref{fMRI_1}). We can see from the heatmap of the pairwise distances of the observations (Fig. \ref{Heatmap_1}) that the existence of a change-point at around $435$ is reasonable. { {In the second sequence (SID-000024), all three methods find the change-point at around $260$, which also appears to be consistent with the heatmap (Fig. \ref{Heatmap_8}).}} Among the three methods, the new test is much more efficient to run (Table \ref{fMRI_1}, last column). We can also observe that here the computation time for $5$-MST is similar to that of ecp as constructing the $5$-MST dominates the overall runtime when $d$ is large.

\begin{table}[h]
\center
\caption{Results of the estimated change-point locations ($\hat\tau$), $p$-values, and the overall runtimes. For the two graph-based methods, the analytical $p$-values are reported; for the ecp method, the $p$-value is based on $999$ permutaions.}
\label{fMRI_1}

\begin{adjustbox}{scale=1.0}
\begin{tabular}{c|lccc}
\hline
 Subject & \multicolumn{1}{c}{Methods} & \c $\hat\tau$ & $p$-value & time cost (minutes) \\ [0.1cm] \hline
 \multirow{3}{*}{\shortstack{ \\ SID-000005}}
 & New (d-a$5$NN) & \c 437 & $<0.001$ & 3.8 \\ [0.1cm]
 & $5$-MST & \c 437 & $<0.001$ & 120.9 \\ [0.1cm]
 & ecp & \c 433 & $0.001$ & 118.4 \\ [0.1cm] \hline
   \multirow{3}{*}{\shortstack{ \\ SID-000024}}
 & New (d-a$5$NN) & \c 260 & $<0.001$ & 3.9 \\ [0.1cm]
 & $5$-MST & \c 260 & $<0.001$ & 147.8 \\ [0.1cm]
 & ecp & \c 261 & $0.001$ & 133.1 \\ [0.1cm] \hline
\end{tabular}
\end{adjustbox}

\end{table}

\subsection{Neuropixels data} \label{Section5.2}

Neuropixels probes are new technology in neuroscience that can record hundreds of sites in the brain simultaneously \cite{Jun2017Nature,CS2019Science}. Here, we analyze a dataset that records the spiking activities of the neurons in the brain of a mouse while it is awake in darkness during spontaneous behavior. The dataset is publicly available at: \texttt{\url{https://figshare.com/articles/dataset/Eight-probe_Neuropixels_recordings_during_spontaneous_behaviors/7739750}}. This dataset contains simultaneous recordings from nine brain regions, with each region having hundreds of recording sequences. 
This dataset was analyzed in \cite{ChenNeuro}, and we follow the same preprocessing procedure there. {{The lengths of all the sequences are the same $n=39,053$, and the dimensions are the numbers of recordings which vary from $d=42$ to $d=334$. We apply the two graph-based methods to all the nine sueqences.}} The results are presented in Table \ref{Neuro_10}. The ecp method is not applicable to this dataset because the memory space is not enough under the same environment as in Table \ref{time_comparison}.

\begin{table}[h]
\center
\caption{Results of the estimated change-point locations ($\hat\tau$), $p$-values, and the overall runtimes (in minutes). For the two graph-based methods, the analytical $p$-values are reported.} 
\label{Neuro_10}

\begin{adjustbox}{scale=1.0}
\begin{tabular}{c|lccc}
\hline
 Region & \multicolumn{1}{c}{Methods} & $\hat\tau$ & $p$-value & time \\ [0.2cm] \hline
  \multirow{2}{*}{\shortstack{ \\ Caudate putamen \\ ($d=176$)}}
 & New (d-a$5$NN) & 35,148 & $<0.001$ & 7.7 \\ [0.1cm]
 & $5$-MST & 35,056 & $<0.001$ & 96.1 \\ [0.2cm] \hline
  \multirow{2}{*}{\shortstack{ \\ Frontal motor \\ ($d=78$)}}
 & New (d-a$5$NN) & 31,081 & $<0.001$ & 6.0 \\ [0.1cm]
 & $5$-MST & 32,242 & $<0.001$ & 77.8 \\ [0.2cm] \hline
  \multirow{2}{*}{\shortstack{ \\ Hippocampus \\ ($d=265$)}}
 & New (d-a$5$NN) & 4,109 & $<0.001$ & 20.7 \\ [0.1cm]
 & $5$-MST & 4,382 & $<0.001$ & 159.1 \\ [0.2cm] \hline
  \multirow{2}{*}{\shortstack{ \\ Lateral septum \\ ($d=122$)}}
 & New (d-a$5$NN) & 29,616 & $<0.001$ & 11.4 \\ [0.1cm]
 & $5$-MST & 29,636 & $<0.001$ & 89.3 \\ [0.2cm] \hline
  \multirow{2}{*}{\shortstack{ \\ Midbrain \\ ($d=127$)}}
 & New (d-a$5$NN) & 20,580 & $<0.001$ & 13.9 \\ [0.1cm]
 & $5$-MST & 20,590 & $<0.001$ & 105.6 \\ [0.2cm] \hline
   \multirow{2}{*}{\shortstack{ \\ Superior colliculus \\ ($d=42$)}}
 & New (d-a$5$NN) & 23,539 & $<0.001$ & 4.0 \\ [0.1cm]
 & $5$-MST & 31,328 & $<0.001$ & 65.4 \\ [0.2cm] \hline
   \multirow{2}{*}{\shortstack{ \\ Somatomotor \\ ($d=91$)}}
 & New (d-a$5$NN) & 30,316 & $<0.001$ & 7.6 \\ [0.1cm]
 & $5$-MST & 30,312 & $<0.001$ & 81.9 \\ [0.2cm] \hline
   \multirow{2}{*}{\shortstack{ \\ Thalamus \\ ($d=227$)}}
 & New (d-a$5$NN) & 28,613 & $<0.001$ & 21.7 \\ [0.1cm]
 & $5$-MST & 28,608 & $<0.001$ & 146.1 \\ [0.2cm] \hline
 \multirow{2}{*}{\shortstack{ \\ V1 \\ ($d=334$)}}
 & New (d-a$5$NN) & 30,226 & $<0.001$ & 17.5 \\ [0.1cm]
 & $5$-MST & 30,338 & $<0.001$ & 173.8 \\ [0.2cm] \hline
\end{tabular}
\end{adjustbox}

\end{table}

We see that the new method on the directed approximate $5$-NN graph is { {on average}} ten times faster than the method in \cite{Chu2019} on $5$-MST. 
Such improvement can be very imperative especially when analyzing large datasets.


\section{Conclusion}

As we enter the era of big data, the importance of the scalibility of statistical methods or data analysis teachniques cannot be overemphasized. Nowadays we are collecting data with exploding sizes (either the dimensionality gets higher or the sequence of observations gets longer). To address this problem, we propose a new nonparametric framework for change-point analysis using the information of approximate $k$-NN graphs. The time complexity of performing our proposed test is $O\left(dn(\log n+k\log d)+nk^2\right)$, and our method is so far the fastest change-point detection method available with a proper control on the false discovery rate.

In constructing the test statistic, we take into account a pattern caused by the curse of dimensionality. As a result, the new test can detect various types of changes (such as change in mean and/or variance, and change in the covariance structure) in long sequences of moderate- to high- dimensional data. Moreover, our method does not impose any distributional assumption on the data, making it desirable in many real applications where the distribution could be heavy-tailed and/or skewed, the dimension of the data could be much higher than the number of observations, and the change could be global or in a sparse/dense subset of coordinates.

We apply our method to two large real datasets, the fMRI images and the Neuropixels recordings, with the former having a very large dimension and the latter a very large sample size. Both examples show that our new method improves the computational efficiency upon the existing methods by a significant amount, while its performance remains as reliable as other state-of-the-art methods.

\ifCLASSOPTIONcaptionsoff
  \newpage
\fi



%

\bibliographystyle{IEEEtran}
\bibliography{References}

%








\end{document}